# $MoSi_2N_4$-like crystals – the new family of two-dimensional materials

T. Latychevskaia[1], D. A. Bandurin[2]*, K. S. Novoselov[3]*

[1]*Paul Scherrer Institute, Forschungsstrasse 111, 5232 Villigen, Switzerland*

[2]*Department of Materials Science and Engineering, National University of Singapore, Singapore, 117575, Singapore*

[3]*Institute for Functional Intelligent Materials, National University of Singapore, Building S9, 4 Science Drive 2, Singapore 117544*

**E-mail:* dab@nus.edu.sg; kostya@nus.edu.sg

## Abstract

Recently-synthesised $MoSi_2N_4$ is the first septuple layer two-dimensional material, which doesn't naturally occurs as a layered crystal, and has been obtained with CVD growth. It can be considered as $MoN_2$ crystal (with a crystal structure of $MoS_2$) intercalating $Si_2N_2$ two-dimensional layer (with the structure similar to InSe). Such classification gave rise to the understanding of the electronic properties of the material, but also to the prediction of other members of the family (many dozens of them) as well as to the way to classify those. Whereas the originally-synthesised $MoSi_2N_4$ is a semiconductor, some of the members of the family are also metallic and some even demonstrate magnetic properties. Interestingly, the room-temperature mobility predicted for such crystals can be as high few thousands $cm^2/V{\cdot}s$ (hole mobility typically higher than electron) with some record cases as high as $5\times10^4$ $cm^2/V{\cdot}s$, making these materials strong contenders for future electronic applications. The major interest towards these materials is coming from the septuple layer structure, which allows multiple crystal phases, but also complex compositions, in particular those with broken mirror-reflection symmetry against the layer of metal atoms.

## Contents

# 1 Introduction

For many years there was a dogmatic approach in the field of two-dimensional (2D) materials that such one atom thick objects need to be originally present as layers in a van der Waals crystal. Indeed, a large number of 2D materials have been obtained by micromechanical[1,2] (Novoselov2004Science, Novoselov2005PNAS) or liquid phase exfoliation[3] (Nicolosi2013Science). The most prominent example is graphene obtained by mechanical exfoliation from graphite. Later, other established chemical synthesis methods: chemical vapor deposition (CVD), etching, as well as various epitaxial techniques, were employed to produce such one-atom thick fabrics[4,5] (Bhimanapati2015ACSNano, Cai2018ChemRev). By using etching method, for example, new 2D materials – the 2D transition metal carbides (MXenes) and 2D transition metal borides (MBenes) – have been synthesized[6-12] (Naguib2011AdvMat, Naguib2012ACSNano, Naguib2014AdvMat, Guo2017JMatChemA, Zhang2018JMatSciTech, Alameda2018JAmChemSoc, Zhang2022JMatChemA). MAX materials (M is a transition metal element, A is an element from group 13 or 14 and X is either carbon or nitrogen) are the three-dimensional (3D) precursor to MXenes, they have a layered structure where $M_{n+1}X_n$ layers are separated by A-layers. Because of relatively weak M-A bonds, the A layers are selectively etched away, leaving 2D MXenes nano-sheets. However, until recently, there were no examples of 2D materials without a layered 3D counterpart – it is not even clear how such materials can be created, as neither exfoliation, nor selective etching would work here.

And yet, a few years ago the first 2D material without 3D counterpart was synthesised[13,14] (Hong2020Science, Novoselov2020NatSciRev). Researchers managed to passivate the surface of monolayer $MoN_2$ with silicon, thus quenching the growth at the monolayer stage, and creating a 2D material of a composition $MoSi_2N_4$, with septuple atomic layers arranged as N-Si-N-Mo-N-Si-N, Fig. 1. The complex layered structure as well as the multiple possibilities for substitutions for all three elements give rise to a new family of 2D materials, very rich in compositions as well as in crystal phases. So far only two members of the family have been experimentally synthesised so far ($MoSi_2N_4$ and $WSi_2N_4$)[13] (Hong2020Science), but the overall strategy for the growth of such crystals is clear and there is no doubt that more of the members of the family will be obtained experimentally in the very near future. In the following we will categorize the multiple possible members of this family, and review their most exciting physical and chemical properties.

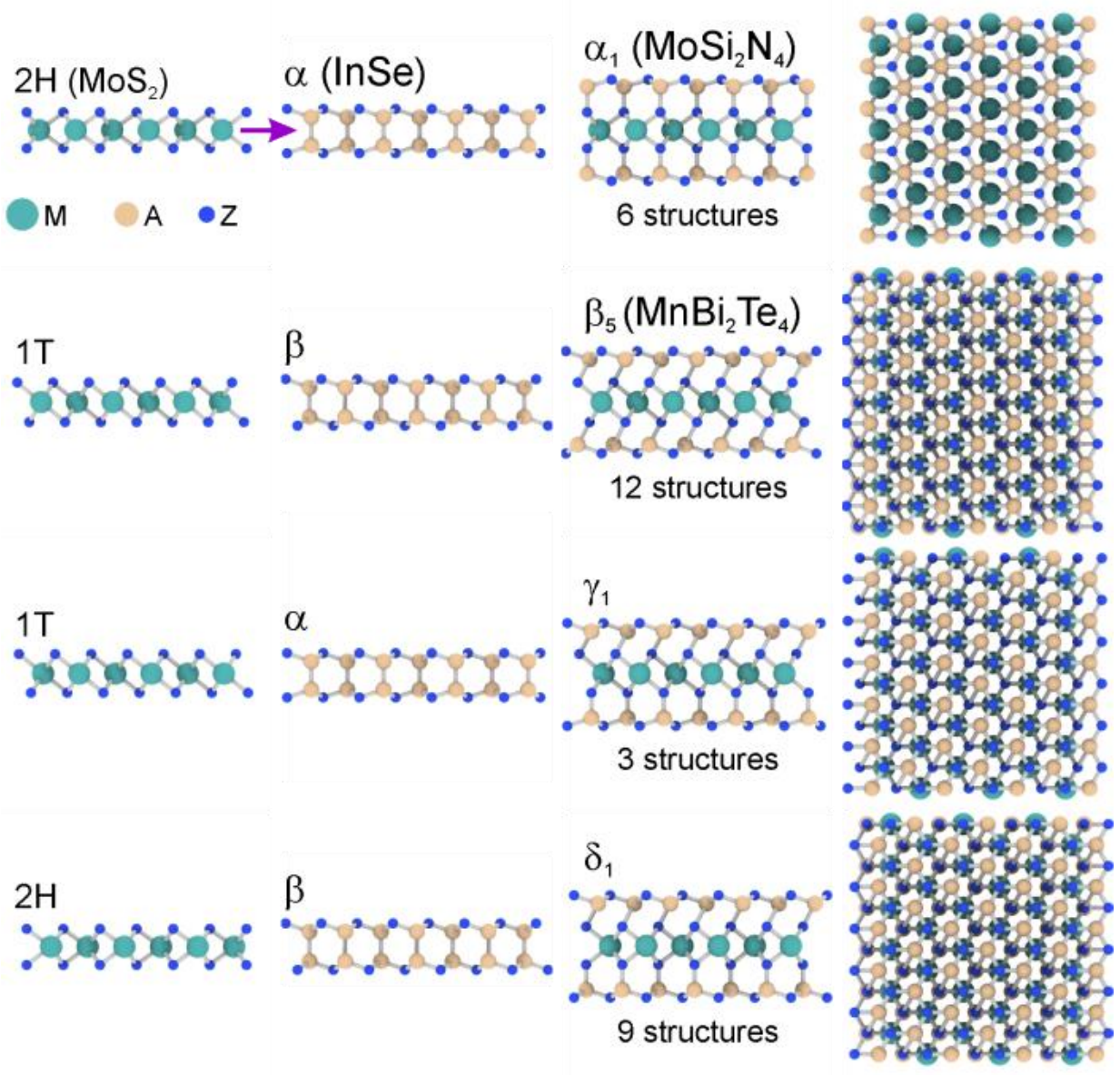

**Fig. 1. Structure of $MoSi_2N_4$ family.** Four symmetry types of $MA_2Z_4$ structures, characterised based on the symmetry of the parent $MoSi_2$ and InSe structures. The structures are drawn using the data provided in ref. 15.

| No. | phase | symmetry group | $a$ / Å |
|---|---|---|---|
| 1 | $\alpha_1$ | P$\underline{6}$m2 | 2.909 |
| 2 | $\alpha_2$ | P$\underline{6}$m2 | 2.901 |
| 3 | $\alpha_3$ | P$\underline{6}$m2 | 2.835 |
| 4 | $\alpha_4$ | $P\bar{6}m2$ | 2.842 |
| 5 | $\alpha_5$ | $P\bar{6}m2$ | 2.864 |
| 6 | $\alpha_6$ | $P\bar{6}m2$ | 2.845 |
| 7 | $\beta_1$ | $P\bar{3}m1$ | 2.909 |
| 8 | $\beta_2$ | $P\bar{3}m1$ | 2.921 |
| 9 | $\beta_3$ | $P\bar{3}m1$ | 2.874 |
| 10 | $\beta_4$ | $P\bar{3}m1$ | 2.857 |
| 11 | $\beta_5$ | $P\bar{3}m1$ | 2.842 |
| 12 | $\beta_6$ | $P\bar{3}m1$ | 2.841 |
| 13 | $\beta_7$ | $P\bar{3}m1$ | 2.900 |
| 14 | $\beta_8$ | $P\bar{3}m1$ | 2.933 |
| 15 | $\beta_9$ | $P\bar{3}m1$ | 2.878 |
| 16 | $\beta_{10}$ | $P\bar{3}m1$ | 2.853 |
| 17 | $\beta_{11}$ | $P\bar{3}m1$ | 2.876 |
| 18 | $\beta_{12}$ | $P\bar{3}m1$ | 2.859 |
| 19 | $\gamma_1$ | P3m1 | 2.882 |
| 20 | $\gamma_2$ | P3m1 | 2.872 |
| 21 | $\gamma_3$ | P3m1 | 2.868 |
| 22 | $\delta_1$ | P3m1 | 2.871 |
| 23 | $\delta_2$ | P3m1 | 2.872 |
| 24 | $\delta_3$ | P3m1 | 2.854 |
| 25 | $\delta_4$ | P3m1 | 2.905 |
| 26 | $\delta_5$ | P3m1 | 2.839 |
| 27 | $\delta_6$ | P3m1 | 2.888 |
| 28 | $\delta_7$ | P3m1 | 2.854 |
| 29 | $\delta_8$ | P3m1 | 2.841 |
| 30 | $\delta_9$ | P3m1 | 2.875 |

**Table 1. Structural properties of 30 $MA_2Z_4$ monolayer candidates.** *a* is the in-plane lattice constant. *xy* represent the x and y coordinates of the atomic positions in the order of N1-N2-N3-N4-Mo1-Si1-Si2. a, b and c represent the sites in trigonal lattice. $z_{N1}$, $z_{N2}$, $z_{N3}$, $z_{N4}$, $z_{Mo1}$, $z_{Si1}$ and $z_{Si2}$ are the factional *z*- coordinate the atomic positions; the *z*-axis unit cell is 30 Å. Adapted with permission from ref. 15.

# 2 Synthesis and experimental characterization

In order to synthesize $MoSi_2N_4$ crystals, a sandwich of Mo and Cu foils was exposed to $NH_3$ gas as nitrogen source, and a pure silicon (or quartz) plate was placed either upstream or directly above the copper foil as a silicon source. The growth temperature was 1080°C. In the absence of the silicon supply – only nanocrystals of non-layered $Mo_2N$ of approximately 10 nm in thickness were grown. Addition of silicon resulted in fairly large, triangular domains of $MoSi_2N_4$, which could fuse into a continuous film upon further growth. TEM analysis confirmed the septuple layer structure.

# 3 Universal structure of the family

## 3.1 Symmetry consideration for $MA_2Z_4$

The first septuple layer 2D material – $MnBi_2Te_4$ – was synthesized in 2013[16] (Lee2013 Crystengcomm) and later intensively studied for its magnetic and topological properties[17-23]

(Gong2015ChinPhysLett, Peng2019PRB, Li2019SciAdv, Zhang2019PRL, Otrokov2019PRL, Deng2020Science, Gao2021Nature). However, only with the appearance of the $MoSi_2N_4$ it was realized that such crystals compose a whole new class of materials. In order to classify the possible structures, the easiest way is to think about such materials not as $MoN_2$ layer passivated with silicon (even though this way of thinking helps when studying the synthesis process), but rather as a $MoN_2$ layer (with a structure similar to that of, for instance $MoS_2$ transition metal dichalcogenide (TMD)) intercalating $Si_2N_2$, where single SiN layer has a structure similar to that of InSe.

In more general terms, a $MA_2Z_4$ monolayer can be constructed by intercalation of a $MZ_2$ layer into a $A_2Z_2$ layer ($MZ_2$ layer is sandwiched between AZ layers, Fig. 1)[15] (Wang2021NatCommun). Here M can stand for elements of transition metal groups IVB, VB and VIB; A – for Si or Ge; and Z – for N, P or As. The further classification originates from the fact, that a $MZ_2$ monolayer can exist in 2H or 1T high symmetry phase and a $A_2Z_2$ monolayer – in $\alpha$ or $\beta$ phases. Thus, an $MA_2Z_4$ 2D crystal structure is given by combination of one of the phases of $MZ_2$ monolayer with one of the phases of $A_2Z_2$ monolayer. Each side of $MZ_2$ has three high-symmetry location sites for A and Z atoms in AZ layer, thus making 6 possible structural arrangements. Because the arrangement of AZ layers on two sides of $MZ_2$ can be different, this gives in total 36 different configurations for each pair of $MZ_2$ and $A_2Z_2$ monolayers. Moreover, there are 144 possible configurations when considering combinations of 2H or 1T phases of $MZ_2$ with either $\alpha$ or $\beta$ phases of $A_2Z_2$. It has been shown[15] (Wang2021NatCommun), that only 39 of those configurations are original – the others are related via symmetry transformations. Only 30 of them are found to be stable (Table 1), and they are split into four groups: α (6 structures), β (12 structures), γ (3 structures) and δ (9 structures), shown in Fig. S1. Calculations demonstrate that it is the $\beta_5$ phase which has the lowest energy[21,22,24,25] (Otrokov2019PRL, Otrokov2019Nature, Deng2020Science, He2020npjQuanMat) for $MnBi_2Te_4$, and $\alpha_1$ phase for $MoSi_2N_4$ and $WSi_2N_4$, as indeed has been observed in the experiments[13] (Hong2020Science).

To classify the $MA_2Z_4$ compounds even further, it is important to consider the number of valence electrons[15] (Wang2021NatCommun). Thus, it is the $\beta_2$ phase which has the lowest energy for crystals with 32 electrons where M belongs to transition metal groups IVB, VB or VIB, A=Si or Ge, and Z is N, P, or As. Different situation is found for the 33 and 34 valence electrons compounds. It is the $\alpha_1$ phase which has the lowest energy for $MSi_2N_4$ (where M is V, Nb, Ta, Cr, Mo, W), $MGe_2N_4$ (for M=Nb, Ta, Mo, W) and $MSi_2P_4$ (M=Nb, Ta – 33 electrons systems); and $\alpha_2$ phase for $MSi_2P_4$ (M=Cr, Mo, W – 34 electron systems), $MGe_2P_4$ (M=V, Nb, Ta, Cr, Mo, W), $MSi_2As_4$ (M=V, Nb, Ta, Mo, W) and $MGe_2As_4$ (M=V, Nb, Ta, Mo, W). The monolayers where the metal atom is either from the alkali-earth group or group IIB are found to be either in $\beta_1$ or $\beta_2$ phase.

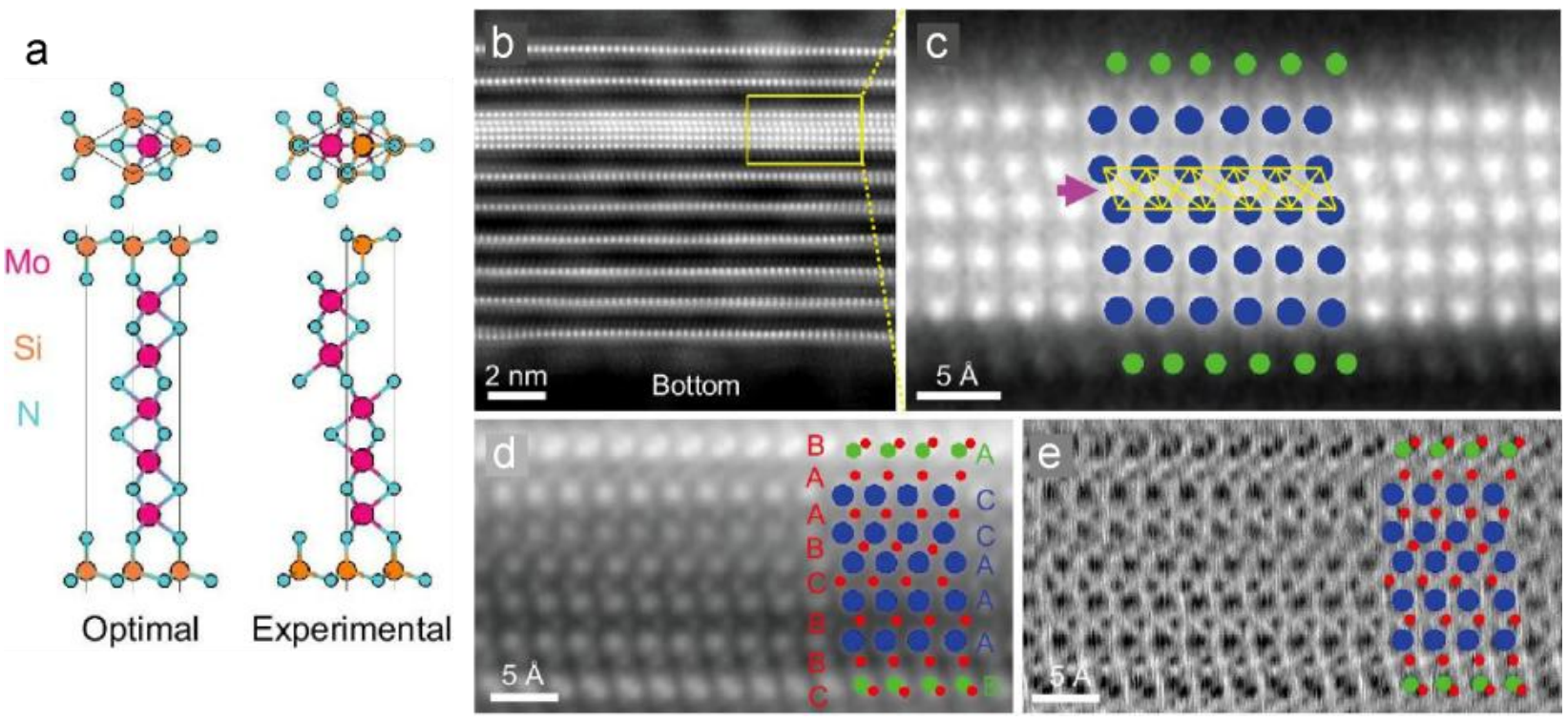

**Fig. 2. Transmission electron microscopy study of $MoSi_2N_4(MoN)_4$. a,** The atomic models of the energetically favourable and experimental structure of $MoSi_2N_4(MoN)_4$. **b** – **c**, high-angle annular dark-field imaging scanning transmission electron microscopy (HAADF-STEM) image of $MoSi_2N_4(MoN)_4$ confined in multilayer $MoSi_2N_4$ (**b**) and the zoomed-in view of the $MoSi_2N_4(MoN)_4$ (**c**). **d** – **e**, integrated differential phase contrast (iDPC) (**d**) and differentiated differential phase contrast (dDPC) (**e**) images of $MoSi_2N_4(MoN)_4$. The blue, green and red balls represent the Mo, Si and N atoms, respectively. Reproduced with permission from ref. [26].

## 3.2 Related structures – Janus crystals

The layer structure of $MA_2Z_4$ materials makes it possible to create the so-called Janus structures that lucks the inversion symmetry with respect to the plane passing through the middle atomic layer (for instance - molybdenum). Such Janus structures have been previously studied both theoretically and experimentally[27] (Zhang2020JMaterChem). Potentially, materials like $MoSiGeN_4$ (as well as many other materials of the family $MABZ_4$) can be synthesized simply by adding not only Si, but also Ge during the growth process. However, the experimental conditions which would lead to elemental separation according to the layers are yet to be explored.

Indeed, many Janus structures of the $MABZ_4$ family have been found[28-30] (Guo2021JMaterChem2464, Yu2021ACS, Guo2021JMtChem7465) stable through the DFT calculations: $MoSiGeN_4$, $WSiGeN_4$, $SrAlGeSe_4$ and others. Interestingly, the calculations indeed confirm separation of the elements between the layers. The peculiar feature of the Janus materials is the possibility of existence of the intrinsic out-of-plane electric field. Such field can induce the Rashba spin splitting around the $\Gamma$ point, as indeed has been seen in the calculations[30] (Guo2021JMatChemC) for both $MoSiGeN_4$, $WSiGeN_4$. This provides a complex spin texture in *x*-*y* plane, with no *z* component of the spin.

## 3.3 Related structures - $MoSi2N_4(MoN)_4$

The intercalation analogy can be taken further. It has been demonstrated[26] (Liu2023NatSciRev) that increase of the flux of $NH_3$ during the growth creates more MoN layers intercalated between SiN layers, Fig. 2. The generic formula of the resulting compound is $MoSi_2N_4(MoN)_{4n}$, where n=0, 1, 2, 3… . Thus, the $MoSi_2N_4(MoN)_4$ compound (n=1) can be seen as 5 layers of MoN ($Mo_5N_6$) sandwiched between two SiN layers. Thicker samples with n up to 10 have also been observed in the experiment. Interestingly, experimentally obtained $MoSi_2N_4(MoN)_4$ crystals demonstrate a Mo octahedral stacking fault, even though theory predicts that the perfect structure (which contains only the MoN triangular prisms) has an energy, which is 108 meV lower that the one with the stacking fault.

Density functional theory (DFT) calculations suggest[26] (Liu2023NatSciRev) that unlike $MoSi_2N_4$ (which is a semiconductor) – the thicker $MoSi_2N_4(MoN)_4$ compound is metallic. Moreover, it has been predicted that it can be a phonon-mediated superconductor with the critical temperature of 9.02 K.

# 4 Mechanical properties

The mechanical properties of $MoSi_2N_4$ crystals have been measured experimentally via atomic force microscopy (AFM) nanoindentation[13] (Hong2020Science). The Young modulus and the breaking strength were found to be 491.4±139.1 GPa and 65.8±18.3 GPa respectively; the numbers confirmed with DFT calculations[31] (Li2022PhysicaE). The origin of such relatively high Young modulus was understood once comparing the mechanical properties of $MoSi_2N_4$ and $MoSi_2N_4(MoN)_{4n}$ compounds. It has been observed that the $MoSi_2N_4$ has much higher Young modulus than crystals that contain $Mo_{n+1}N_{n+2}$ layers, and the Young modulus of $MoSi_2N_4(MoN)_{4n}$ changes only slightly with the number of MoN layers[26] (Liu2023NatSciRev). This suggests that the major contribution to the higher Young modulus is mainly provided by the of SiN layers, which can be explained by ionic nature of Si-N bonds, compared to more covalent Mo-N.

Some of the phases of $MA_2Z_4$ compounds demonstrate the lack of inversion symmetry and thus can possess piezoelectric properties. Thus, the $\alpha$ phases belong to the $P\bar{6}m2$ group (no inversion symmetry), thus exhibiting piezoelectric properties and the $\beta$ phases belong to the $P\bar{3}m1$ group – preserving such symmetry[32] (Guo2021CompMatSci). The piezoelectric properties have been calculated by many authors for a number of different semiconducting $MA_2Z_4$ compounds[30-34] (Guo2021JMaterChem7465, Li2021PhysicaE, Guo2021CompMatSci, Bafekry2021JPhysD, Mortazavi2021NanoEnergy). For $\alpha_1$-$MoSi_2N_4$ and $\alpha_1$-$WSi_2N_4$ (which have been experimentally synthesized) the in-plane piezoelectric strain $d_{11}$ are 1.14 pm/V and 0.78 pm/V respectively[32] (Guo2021CompMatSci). These values are somewhat smaller than those found in 2D TMD semiconductors[35,36] (Blonsky2015ACSNano, Duerloo2012JPhysChem), mainly due to large in-plane stiffness. Among other possible compounds, those based on phosphorous predicted[32] (Guo2021CompMatSci) to have significantly higher $d_{11}$, for instance 4.91 pm/V for $\alpha_1$-$MoSi_2P_4$ and 6.12 pm/V for $\alpha_1$-$CrGe_2P_4$.

Sliding ferroelectricity is an exciting phenomenon, observed in multilayer van der Waals heterostructures[37,38] (Geim2013Nature, Novoselov2016Science) where certain stacking results in local breaking of the vertical inversion symmetry[39-41] (Woods2021NatCommun,

ViznerStern2021Science, Weston2022NatNanotech). Similarly, the relatively low symmetry of $MA_2Z_4$ compounds offer bilayer stacking configurations with strong ferroelectric domains. Thus, vertical polarisations can be[42] (Zhong2021JMatChemA) as high as 3.36 pC/m and 3.05 pC/m for $MoSi_2N_4$ and $MoGe_2N_4$, and 2.49 pC/m and 3.44 pC/m for $CrSi_2N_4$ and $WSi_2N_4$, respectively, which is significantly higher than, for instance for bilayer $WTe_2$ (predicted[43] (Yang2018JPhysChemLett) to be ~0.37pC/m).

# 5 Thermal conductivity and thermoelectric properties

The large in-plane stiffness – being detrimental for piezoelectric properties of $MA_2Z_4$ compounds – can be beneficial for their thermal conductivity. The phonon spectra of such compounds consist of 3 acoustic branches and 18 optical branches[44] (Yin2021ASC) with the acoustic phonons contributing the most to the thermal conductivity (typically of the order of 75%)[34] (Mortazavi2021NanoEnergy). The important parameter determining the thermal conductivity is the phonon group velocity, which by itself depends on the width of the dispersion of a particular phonon band (the wider the band – the higher the group velocity). It turns out that the width of the phonon bands depends strongly on the A and Z atoms, but only marginally on the M atom[34] (Mortazavi2021NanoEnergy). Thus, it has been predicted that room temperature thermal conductivity for $MoSi_2N_4$ and $WSi_2N_4$ are 439 W/m·K and 503 W/m·K respectively (though lower values can also be found in literature[45,46] (Yu2021NewJPhys, Zhang2022JSolidState)), but drops by a factor of 4 if nitrogen is replaced with phosphorous. Thermal conductivity is found to decay with temperature as 1/T, suggesting the Umklapp phonon scattering as the dominant scattering mechanism[45,46] (Yu2021NewJPhys, Zhang2022JSolidState). Interestingly, thermal conductivity is not sensitive for the choice of the M atom if they belong to the same group[44] (Yin2021ASC). At the same time, as the bond-strength depends on the valence of the M atom – the thermal conductivity changes significantly depending on to which group in the periodic table the M atom belongs to.

The high thermal conductivity would reduce the efficiency of the thermoelectric effect in such materials. The figure of merit for $MoSi_2N_4$ at room temperatures has been predicted to be of the order of 0.05, though increasing with temperature due to decrease of the thermal conductivity.

# 6 Band structure

The band structure of $MA_2Z_4$ compounds strongly depends on the number of valence electrons. Thus, $MA_2N_4$ compounds (nitrides) with 32 electrons are found to be semiconductors[15] (Wang2021NatCommun). At the same time, $MA_2P_4$ compounds (phosphides) and $MA_2As_4$ (arsenides) with the same number of electrons predicted to be metallic. This is due to the fact that the hybridization between the M atom and phosphorous or arsenic is much weaker than between M atom and nitrogen.

Most of the 34 electron systems are semiconductors (which indeed has been experimentally confirmed for $MoSi_2N_4$, which has been confirmed as a semiconductor[13] (Hong2020Science)), and only $CrGe_2N_4$, $CrSi_2As_4$, $CrGe_2As_4$ are predicted to be metallic ferromagnets[15] (Wang2021NatCommun). Many of the systems with 33 electrons exhibit ferromagnetic properties, and those which don't are typically metallic, since the unpaired electron would give rise to a half-filled band.

# 7 Magnetic Properties

Among the variety of 2D materials, only a few exhibit magnetism in a 2D limit. The emergent family of $MA_2Z_4$ materials promises to change this situation as they were predicted to feature diverse, intrinsic magnetic properties even in the monolayer form. The freedom of arranging three different elements in a seven-layered crystal structure makes $MA_2Z4$ materials rich with various physical properties, including magnetism. Generally, the M atoms of $MA_2Z_4$ nano-sheets (M = V, Nb, or Ta) are responsible for the ferromagnetic states as the magnetic moments are found to be localized on the d orbital of M-atoms[47,48] (Zhao2023ApplSurfSci, Chen2021ChemA). When M is either V, Nb, or Ta atoms, $MA_2Z_4$ materials were predicted to exhibit a magnetic ground state and can be either ferromagnetic or antiferromagnetic in nature. V-based monolayer $VA_2Z_4$ materials are semiconducting ferromagnets whereas $VGe_2N_4$ are ferromagnetic half-metals with a magnetic moment of 1 $\mu_B$ per metal atom (V)[48,49] (Chen2021ChemA, Li2022NewJPhys). For example, the ground states of $VSi_2Z_4$, $VSi_2P_4$ and $VSi_2N_4$ are ferromagnetic; these materials exhibit spin-gapless semiconducting properties, and can offer spin-filtering[50] (Feng2022APL). Magnetism, similarly to other properties of $MA_2Z_4$ materials, has been also predicted to be sensitive to strain. In particular, numerous phase changes are envisioned in $MA_2Z_4$ crystals: from compression-enabled suppression of magnetism to semiconductor-to-half-metal transitions under biaxial strain[47,50] (Zhao2023ApplSurfSci, Feng2022APL). The presence of magnetic order in this new class of 2D materials may find its applications in the field of spintronics.

The magnetic moments in $NbSi_2N_4$, $VSi_2N_4$, and $VSi_2P_4$ are theoretically found to be aligned in-plane and, therefore, have easy magnetization with negligible magnetic anisotropy[47,51] (Zhao2023ApplSurfSci, Akanda2021APL). The magnetic anisotropy energy is dependent on the metal atoms. An increase in strain in the $ZnSi_2N_4$ layer shows that it can change the magnetic anisotropy from perpendicular to in-plane magnetic anisotropy[47] (Zhao2023ApplSurfSci). These materials also display high Curie temperatures[49,50] (Feng2022APL, Li2022NewJPhys). For $VSi_2X_4$ nanosheets, the Curie temperature is in the range of 230–250 K. For $VGe_2N_4$, it is reported that the ferromagnetic phase is stable under strain (tensile or compressive), and the Curie temperature can reach 325 K, a value over room temperature, by the presence of compressive strain. The Curie temperature of $VSi_2N_4$ is estimated to be around 507 K[51] (Akanda2021APL). Overall, the value of Curie temperature is higher compared to the other 2D magnets like $CrI_3$ (45 K)[52] (Huang2017Nature) or $Fe_3GeTe_2$ (130 K)[53] (Fei2018NatMater). The substitutional doping of transition metals can introduce alternate spin channels and induce magnetic moments in nonmagnetic $MA_2Z_4$ materials such as $MoSi_2N_4$. Mn-doped $MoSi_2N_4$ is reported to exhibit spin-polarization up to 86% and has a potential for applications in spin-filtering[54] (Abelati2022PhysChemChemPhys). A nonmagnetic $MA_2Z_4$ state can be transformed into a magnetic one by doping of this kind at different atomic sites (M, A, or Z)[54-56] (Abelati2022PhysChemChemPhys, Ding2022ApplSurfSci, Ray2021ACSomega). It can be easily perceived that by replacing metal atoms by means of substitution, doping, or alloying and by applying external strain, the magnetic behaviour of the $MA_2Z_4$ family of materials can be regulated to a great extent and engineered according to a targeted application. The freedom of substitution of M, A, or Z site atoms gives rise to different magnetic moments, spin channels, and conductivity suitable for the exploration of both fundamental physics and spintronic applications.

# 8 Valley properties

## 8.1 Berryology and valley-contrasting phenomena in $MA_2Z_4$ family

The Berry phase of the electronic wave function, along with associated physical quantities such as Berry connection and Berry curvature, can impact material properties in a plethora of spectacular and sometimes counterintuitive ways[57] (Xiao2010RevModPhys). It is responsible for a broad range of phenomena, from various types of (quantum) Hall effects and orbital magnetism to ferroelectricity and valley-contrasting physics, Fig. 3. The latter has particularly important implications for the development of advanced electronic and optoelectronic devices with enhanced performance and novel functionalities. Valley-dependent physical effects may manifest themselves in materials characterized by several local extrema (termed valleys) in the energy dispersion. In such materials, the valley degree of freedom (or valley index) can be controlled and manipulated similarly to spin degree freedom and can transfer information, encoded in its value, across the material. Valleytronics thus may offer an interesting alternative to spintronics, as the valley index is more robust with respect to smooth disorder, low-energy phonons, and mild deformations. Furthermore, spin-orbit coupling (SOC) in some multi-valley materials may promote a strong spin-valley locking, so that any impact applied to the spin degree of freedom will manipulate the valley index and vice versa paving the way for an interplay between spintronics and valleytronics studies.

## 8.2 Spin-valley splitting

Some representatives of the $MA_2Z_4$ family have been predicted to serve as an interesting platform for valleytronics and offer a convenient and versatile alternative to extensively studied transition metal dichalcogenides (TMDC) compounds with $MX_2$ composition (where M stands for a transition metal and X stands for chalcogen atoms). Some $MA_2Z_4$ materials (e.g., $MoSi_2N_4$, $WSi_2N_4$, and $MoSi_2As_4$) are characterized by two degenerate but not equivalent valleys located at the K and K' high symmetry points of the hexagonal Brillouin zone, connected by time reversal symmetry[58] (Li2020PRB). Strong spin-orbit coupling and the broken inversion symmetry, predicted in most of these materials, may result in sizable spin splitting and the emergence of non-zero Berry curvature in the conduction and valence bands, opposite for the two valleys. In some single-layer representatives of the $MA_2Z_4$ family, such as $CrSi_2N_4$ and $CrSi_2P_4$, the valley spin splitting is particularly pronounced and can be as large as 0.13 and 0.17 eV for $CrSi_2N_4$ and $CrSi_2P_4$, respectively[59] (Liu2021JPhysChemLett). Importantly, one of the advantages of $MA_2Z_4$ over TMDC $MX_2$ compounds in the context of valleytronics is a much larger span of available band gaps, including the technologically important mid-infrared spectrum[60] (Yuan2022PRB), which until now has only been addressed by highly biased bilayer graphene, black phosphorus, and black arsenic.

## 8.3 Valley-contrasting effects

$MA_2Z_4$ materials allow for effective steering of the valley degree of freedom using transport, optical, and magnetic means. For example, strong spin splitting in $CrSi_2N_4$ and $CrSi_2P_4$ monolayers may give rise to coexisting spin and valley Hall effects[59] (Liu2021JPhysChemLett), i.e. the emergence of valley and spin polarization perpendicularly to the applied electric field. Furthermore, valley-contrasting optical circular dichroism has been predicted to emerge in these materials. However, in contrast to TMDC $MX_2$ compounds, multiple-folded valleys, predicted to emerge in $MoSi_2As_4$ monolayer, enable resonant excitations of non-degenerate spin states hosted by one valley using circularly polarized light of opposite chiralities[61] (Yang2021PRB). Moreover, a monolayer of $VSi_2N_4$ is a ferromagnetic semiconductor in which broken time reversal symmetry can lift the degeneracy between K and K' valleys[62] (Cui2021PRB). By tuning magnetization orientation from in-plane to out-of-plane, a large valley polarization can be generated, that has been predicted to result in the anomalous valley Hall effect. Strain is another intriguing means of controlling valley polarization. Only 4% of tensile strain can cause a spontaneous polarization of $VSi_2N_4$ monolayer whereas biaxial strain, electric field and correlations may result in the valley counterpart of the quantum anomalous Hall effect[63] (Zhou2021npjCompMat).

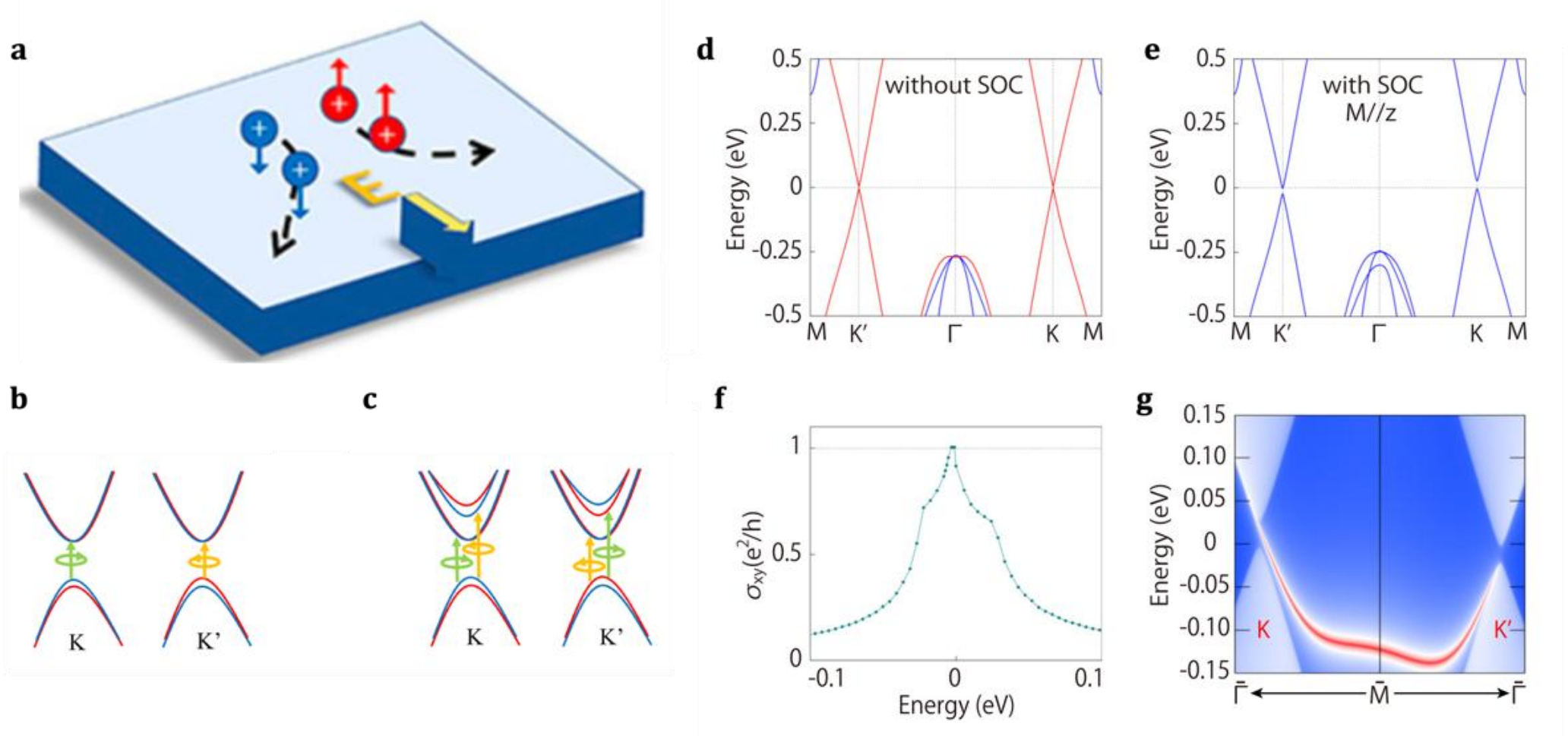

**Fig. 3. Valley-contrasting properties of the $MA_2Z_4$ family. a,** Diagram of the spin and valley Hall effects in hole-doped single layer $CrSi_2X_4$; adapted with permission from ref. [59]. **b,** Conventional valleys and **c,** multiple-folded valleys. Circles with arrows indicate the chirality of light, and arrow length denotes the photon energy. Red and blue represents up- and down-spin states, respectively; adapted with permission from ref. [61]. **d – e,** Band structure of monolayer $VSi_2P_4$ with (**e**) and without (**d**) Spin-orbit coupling (SOC) (the red and blue colors indicate spin-up and spin-down channels, respectively). **f**, Anomalous Hall conductivity $\sigma_{xy}$ versus energy for the case shown in (**e**). **g**, shows the corresponding edge spectrum for the quantum anomalous Hall state in (**e**). d – g, Adapted with permission from ref. [64].

# 9 Transport Properties

Generally, a rather high intrinsic mobilities at room temperature (for instance 270 $cm^2/V\cdot s$ for electrons 1'200 $cm^2/V\cdot s$ for holes for $MoSi_2N_4$ compound[13] (Hong2020Science)) have been predicted for semiconducting $MA_2Z_4$ compounds[13,15,46,65] (Hong2020Science, Wang2021NatCommun, Zhang2022JSolidStateChem, Li2022JAP). Hole mobility is expected to be generally higher than that for electrons, which is partially because of the reduced effective mass (spin-orbit coupling being one of the reasons)[66,67] (Qiu2022JMatChem, Mortazavi2021MatTodayEnergy). Hole mobilities exceeding $10^5$ $cm^2/V\cdot s$ have been predicted for $\alpha_2$-$WSi_2Sb_4$.

Experiments indeed confirmed the semiconducting nature of $MoSi_2N_4$, though the mobilities observed are significantly lower than predicted[13] (Hong2020Science). Synthesized $MoSi_2N_4$ films appear to be p-doped.

Interestingly, the Janus systems demonstrate[28,68,69] (Guo2021JMaterChem, Lv2022APL, Ding2023EPL) the opposite trend than symmetric compounds: the electron mobilities are significantly higher than the hole mobilities (few thousands $cm^{-2}/V\cdot s$ and few hundred $cm^{-2}/V\cdot s$ for holes). Moreover, bilayer Janus systems can demonstrate[68] (Lv2022APL) even an order of magnitude higher mobilities – up to 58'522.3 $cm^2/V\cdot s$ for electrons. This is due to the increase of the elastic moduli and a decrease in the deformation potential.

The performance of field effect transistors based on $MoSi_2N_4$ has been simulated[70,71] (Sun2021JMaterChem, Ghobadi2022IEEE) both for the high performance and low power applications. Both n- and p-type devices have been shown to satisfy the requirements for the high-performance applications, and the p-type transistors satisfying those for the low power ones[70] (Sun2021JMaterChem). This is because the effective mass for holes falls into the sweet range: it is high enough to guarantee a significant density of states, but low enough to secure high mobility of carriers.

Electrical contacts are extremely important for such devices. $MA_2Z_4$ form Schottky or Ohmic type contact with metals. The probability of a particle tunnelling through Schottky barrier in the formed structure is given by

$$P_{\mathrm{TB}}\left(w_{\mathrm{TB}},\Phi_{\mathrm{TB}}\right)=\exp\left(-2w_{\mathrm{TB}}\frac{\sqrt{2m_e\Phi_{\mathrm{TB}}}}{\hbar}\right)=\exp\left(-2w_{\mathrm{TB}}k_{\mathrm{TB}}\right), \quad (1)$$

where $\Phi_{\mathrm{TB}}$ and $w_{\mathrm{TB}}$ are the height and width of the tunnelling barrier, respectively that can be estimated from the simulated charge density difference and effective electrostatic potential[72] (Shu2023AdvElMat); $k_{\mathrm{TB}}=\sqrt{2m_e\Phi_{\mathrm{TB}}}/\hbar$ is the wavenumber, $m_e$ is the free-electron mass and $\hbar$ is the reduced Planck's constant. The tunnelling specific resistivity is calculated by using the expression of current density in a tunnel junction at intermediate voltages[73,74] (Simmons1963JAP_p1793, Matthews2018JAP):

$$\rho_{\mathrm{TB}}\left(w_{\mathrm{TB}},\Phi_{\mathrm{TB}}\right)=\frac{4\pi^2\hbar w_{\mathrm{TB}}^2}{e^2}\frac{\exp\left(2w_{\mathrm{TB}}k_{\mathrm{TB}}\right)}{w_{\mathrm{TB}}k_{\mathrm{TB}}-1} \quad (2)$$

where $e$ is the electron charge. The electronic properties of $MA_2Z_4$ – metal structures are summarized in Table S1 (Schottky barrier heights for electrons and holes, tunnelling barrier height and width, the tunnelling probability and resistivity, lattice mismatch and interlayer distances); the type of $MA_2Z_4$-metal structure as a function of (a) vertical stress (interlayer distance), (b) biaxial strain and (c) electric field are shown in Fig. S2.

Two[75,76] (Wang2021npj2Dmater, Meng2022PRA) and one[76] (Meng2022PRA)-dimensional metal contacts and well as van der Waals contacts[77,78] (Cao2021APL, Binh2021JPhysChemLett) have been investigated. The most noticeable feature is the significantly reduce pinning of the Fermi level, because the states at both the conductance band minimum and valence band maximum are mainly localized at the central metal layer (molybdenum in the case of $MoSi_2N_4$). This guarantees very efficient tunability of the Schottky barrier height depending on the metal utilized for the contacts[75,79] (Wang2021npj2DMater, Ai2023ACS_ApplElMat). Thus, Ti and Pt (as well as graphene) have been suggested among the materials suitable for good metal contacts for n- and p-type devices respectively[75] (Wang2021npj2DMater), Fig. 4. Experimentally, standard Ti/Au contacts have shown to demonstrate good Ohmic behaviour down to nitrogen temperatures[13] (Hong2020Science).

Vertical devices have also been investigated. Thus, it has been shown that $MoSi_2N_4$ can be used as a barrier for magnetic tunnel junctions in combination with $CrI_3$ ferromagnetic contacts[80] (Liu2022PRB). Tunnelling magneto-resistance can reach $10^5$ % due to specific arrangements between the majority/minority bands in $CrI_3$ and the bands in $MoSi_2N_4$.

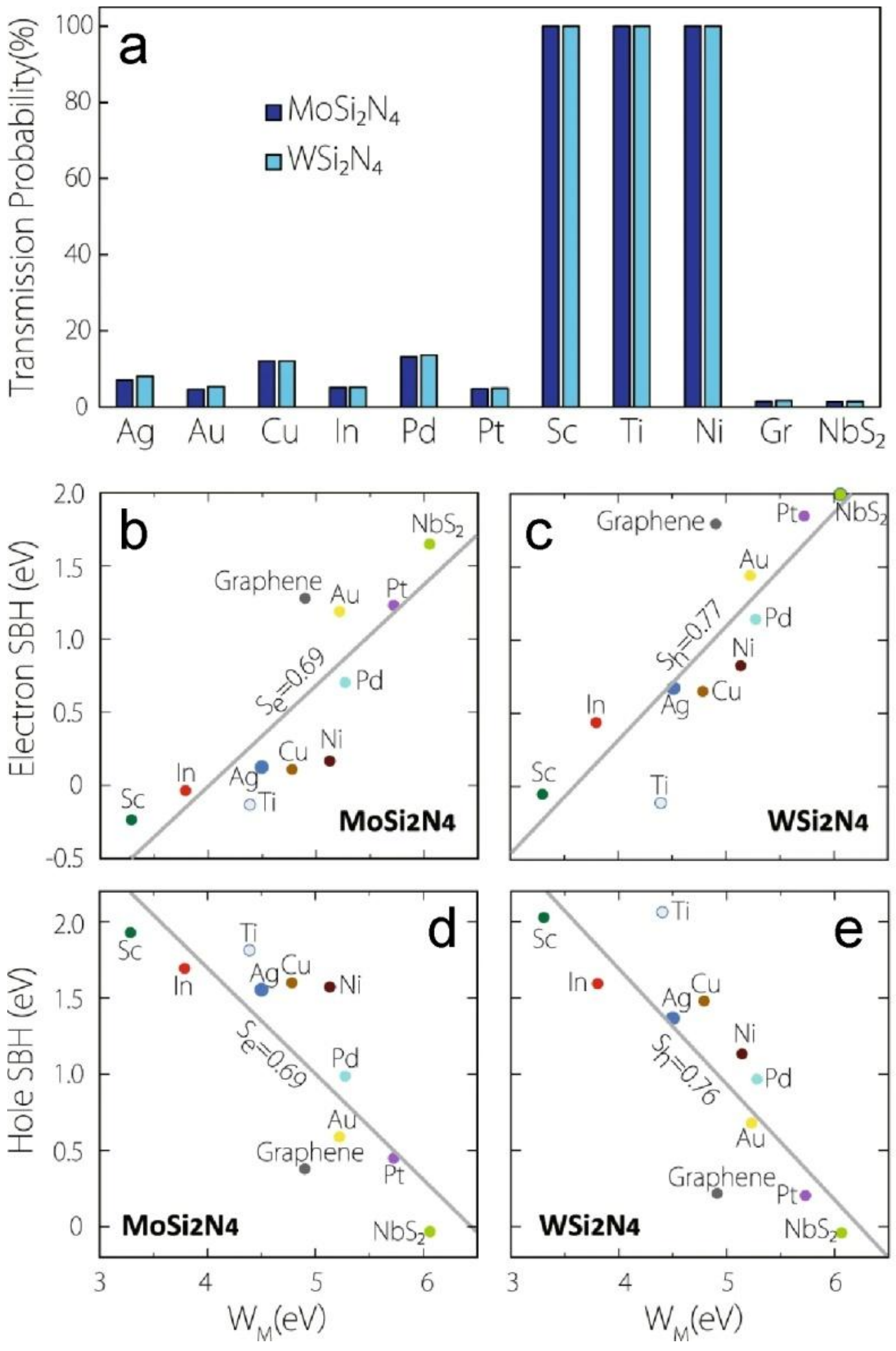

**Fig. 4. Contacts to $MA_2Z_4$ structures. a,** Transmission probability of various metal/$MoSi_2N_4$ and metal/$WSi_2N_4$ heterostructures. **b – c**, show the Schottky–Mott plot for the electron Schottky barrier height of $MoSi_2N_4$ and $WSi_2N_4$, respectively. **d – e**, same as **b – c**, respectively, but for the hole Schottky barrier height. Reproduced with permission from ref. 75.

## 10 Topological properties

Given to a complex composition one can certainly expect that some of the members of the family indeed possess non-trivial topological properties. Indeed, it has been demonstrated[15,81] (Wang2021NatCommun, Wang2022PRB) that a number of $MA_2Z_4$ compounds demonstrate inverted energy bands between W$-d_{z^2}$ and N$-p_z$ orbitals at the $\Gamma$ point of the Brillouin zone. Unfortunately, all of those have been identified as topologically trivial insulator based on the zero value of Chern number. At the same time, it has been demonstrated[81] (Wang2022PRB) some $MA_2Z_4$ compounds can support the so called obstructed Wannier charge centres[82] (Xu2021arXiv), when a portion of a valence electron occupies empty Wyckoff positions, with no atoms sitting at that space (materials with such properties are known as obstructed atomic insulators). In total 16 $MA_2Z_4$ crystals have been identified as obstructed atomic insulators.

What is most promising is that if a cleavage in a crystal passes through such obstructed Wannier charge centres – metallic edge states can be expected, which has indeed been observed[81] (Wang2022PRB) for $(1\bar{1})$ direction boundary of $MoSi_2N_4$ in the $\alpha_2$ phase. Additionally, mid-gap corner states have been observed for both $\alpha_1$ and $\alpha_2$ phases of $MoSi_2N_4$. Such metallic edges and mid-gap states open intriguing opportunity for quantum transport.

# 11 Optical properties

The diverse family of 2D $MA_2Z_4$ includes semiconductors with band gaps ranging from infrared to ultraviolet ranges. Overall, the band gap of the $MA_2Z_4$ monolayer can vary between 0.3–2.57 eV, making it drastically different from conventional transition metal dichalcogenide (TMD) crystals whose band gap range falls into the visible/near-IR range[34] (Mortazavi2021NanoEnergy). The substitution of any of the three elements (M/A/Z) with an element from a similar category or alloying can alter the band structure analogous to the TMDs[83] (Gong2014NanoLetters). For example, the substitution of N-atoms by As- or P-atoms affects the band gap and changes the nature of the nearest interband transition. The onset of optical absorption of $MoSi_2N_4$ is predicted in the visible region (~2.31 eV), whereas that of for $MoSi_2As_4$ falls in the near-infrared range (0.71 eV)[84] (Sun2022npj2DMater). Furthermore, band structure calculations show that monolayers $MA_2N_4$ (e.g., $MoSi_2N_4$) generally exhibit an indirect band gap, whereas $MA_2As_4$ materials (e.g., $MoSi_2As_4$ or $MoSi_2P_4$) are expected to be direct band gap semiconductors[84] (Sun2022npj2DMater). Generally, heavier M-atoms (e.g., W compared to Mo) can induce a pronounced red shift in the band gap compared to the lighter ones[34,58,84-86] (Mortazavi2021NanoEnergy, Sun2022npj2DMater, Yang2022PRB, Li2022Nanomaterials, Li2020PRB). Finally, the band structure can be altered, and light absorption capacity can be increased by applying strain to the layers and employing surface functionalization[87,88] (Zhang2023RSCadv, Liu2023PhysChemChemPhys). The described tunability of the band structure in the 2D $MA_2Z_4$ family makes it promising for polarization-sensitive broadband optoelectronic devices.

Excitons dominate the optical properties in 2D $MA_2Z_4$ as they have high binding energy, similarly to TMD. These materials exhibit excitonic (A & B) absorption features related to spin-orbit coupling-induced band splitting similar to that observed in 2D TMDs and display analogous selection rules for polarized light corresponding to different valleys[13,84,89] (Hong2020Science, Sun2022npj2DMaterAppl, Selig2019PhysRevRes). Higher energy direct interband optical transitions (exciton 'C') with prominent absorption feature is also predicted in layered $MA_2Z_4$ (Fig. 5)[84,90] (Sun2022npj2DMater, Aleithan2016PRB). Similarly to TMDCs, excitons in these materials exhibit high exciton binding energy up to a few hundred meV, which makes them stable against thermal dissociation at room temperature and suitable for excitonic device applications[13,84,89,91,92] (Hong2020Science, Sun2022npj2DMater, Selig2019PhysRevRes, Ai2021PhysChemChemPhys, Wozniak2023Small). However, the presence of the outer SiX layers affects exciton binding energies making them smaller than those generally found in TMDC compounds[84]. The lifetime of excitons $\tau_{\mathrm{X}}$ in 2D semiconductors is proportional to $M_{\mathrm{X}}$ – the effective mass of the exciton, and inversely proportional to the square of $E_{\mathrm{X}}$ – the energy of exciton at zero momentum: $\tau_{\mathrm{X}} \propto M_{\mathrm{X}} / E_{\mathrm{X}}^2$ [93-95](Palummo2015NanoLett, Chen2018NanoLett, Chen2019PRB). As shown in ref.[84] (Sun2022npj2DMater), $\tau_{\mathrm{X}}$ in $MoSi_2N_4$ are comparable to that found in most common TMDs (e.g., $MoS_2$, $MoSe_2$, and $WSe_2$) and is of the order of 1-7 ns at liquid helium temperatures and around 1 ns at room temperature. In contrast, the values for $MoSi_2P_4$ and $MoSi_2As_4$ are approximately two orders of magnitude greater. For example, record high values of $\tau_{\mathrm{X}}$ reaching 370 and 28 ns were predicted in $MoSi_2As_4$ at 4 K and 300 K respectively[84] (Sun2022npj2DMater). This difference was associated with the low exciton energy and the elevated effective mass of the excitons. Besides, the absence of nearby valleys around the band extrema can reduce non-radiative pathways in $MoSi_2Z_4$ nano-sheets, favouring a higher rate of radiative recombination of photo-generated electrons and holes with a higher quantum yield of luminescence[84] (Sun2022npj2DMater). Last, we envision that heterostructures composed of various $MA_2Z_4$ monolayers with aligned or twisted crystallographic axes[96] (Zhong2023PRB) may offer a whole new class of interaction-driven phenomena and give rise to rich phase diagrams[97] (Mak2022NatNanotech).

We also note that, recently, new exciton states have been discovered in 2D semiconductors termed quadrupolar and every-other-layer excitons[98] (Du2024NatureRevPhys). These excitons are characterized by an electrically-tunable dipole moment and feature a many-body phase diagram. Considering that the strong exciton-phonon coupling[99] (Huang2022AdvOptMat) and exciton self-trapping effect[100] (Huang2023AdvOptMat) were reported for $MoSi_2N_4$, we envision that interesting forms of these new tunable exciton states can be also observed in single- few-layer $MA_2Z_4$ sheets.

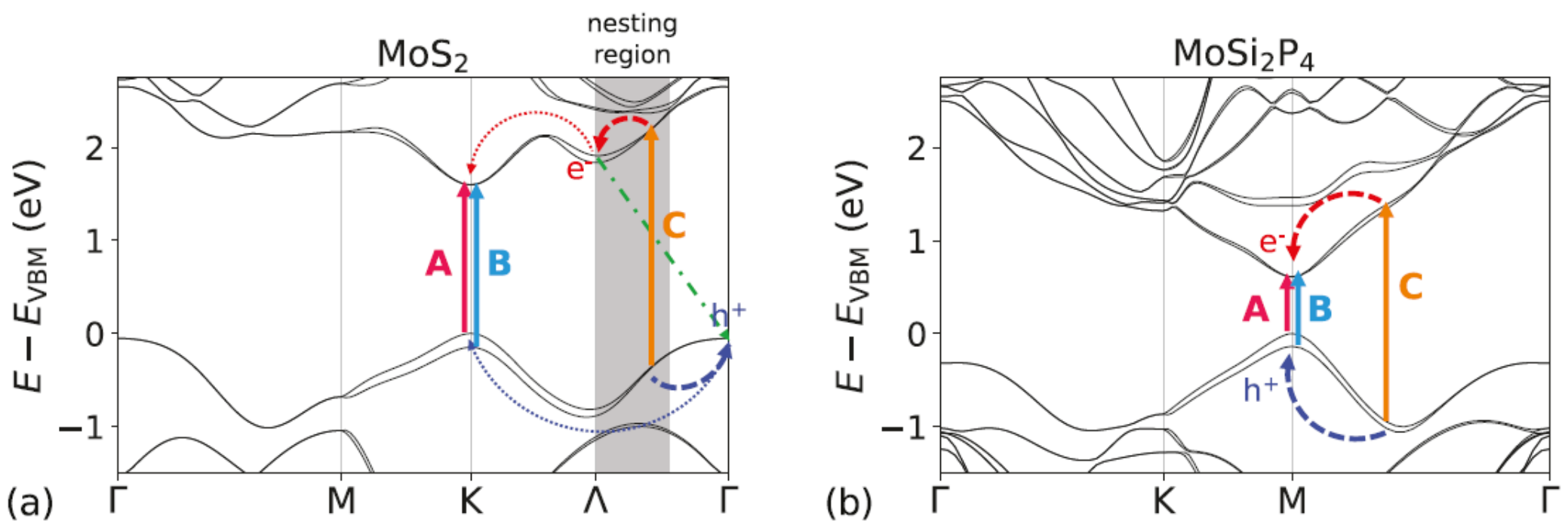

**Fig. 5** Different excitons in (**a**) $MoS_2$ and (**b**) $MoSi_2P_4$ materials and the typical relaxation processes. Adapted with permission from ref. 84.

## 12 Heterostructures

The alignment of the energy bands in $MA_2Z_4$-semiconductor heterostructures can be of normal (I), staggered (II) and broken gap (III) type (Fig. 6a). Liu et al simulated band structures of six perovskites $A_3BX_9$ (A = Cs; B = In, Sb, Bi; and X = Cl, Br, I) and six $MA_2N_4$ (M = Cr, Mo, Ti; A = Si; and Z = N, P) materials and showed that $A_3B_2X_9/MA_2Z_4$ heterostructures can exhibit all three types of band alignment (I, II and III)[101] (Liu2022JApplPhys). Type II alignment is the preferred band alignment, for it allows for an effective electrons and holes separation, avoiding their recombination and thus extending the lifetimes of these charge carriers. External vertical or biaxial strain, or electric field cause charge redistribution within and between the layers, which in turn changes the bands distribution of the overall structure. Thus, these external forces can tune the band alignment type (I, II or III), the band gap value, and the band gap type (direct, indirect)[102-111] (Fang2022PhysicaE, Nguyen2022PRB, Ng2022APL, Xu2023JPhysChemC, Cai2021JMatChemC, Wu2021APL, Pei2023PhysicE, Nguyen2021JPhysChemLett, Wang2021Nanomaterials, Zhang2023JAlloysComp). The applied strain is typically varied in the range of ±10%, and the applied electric field within the range ±0.5 V/Å. The bangap typically decreases under applied strain or electric field, ultimately reaching zero. As an example, $MoSi_2N_4/MoS_2$ structure changing its electronic properties under applied biaxial strain and electric field is shown in Fig. 6b-f[105] (Xu2023JPhysChemC) (Fig. 6b-f).

Various heterostructures created by combining $MA_2Z_4$ layer combined with semiconductors, metals, dielectrics, and laterally stitched structures were recently overviewed by Tho et al.[112] (Tho2023_ApplPhysRev). The electronic properties of $MA_2Z_4$-based vdWHs are summarized in TableS2; the band alignment and bandgap type (direct, indirect) as a function of (a) vertical stress (interlayer distance), (b) biaxial strain and (c) electric field are shown in Fig. S3.

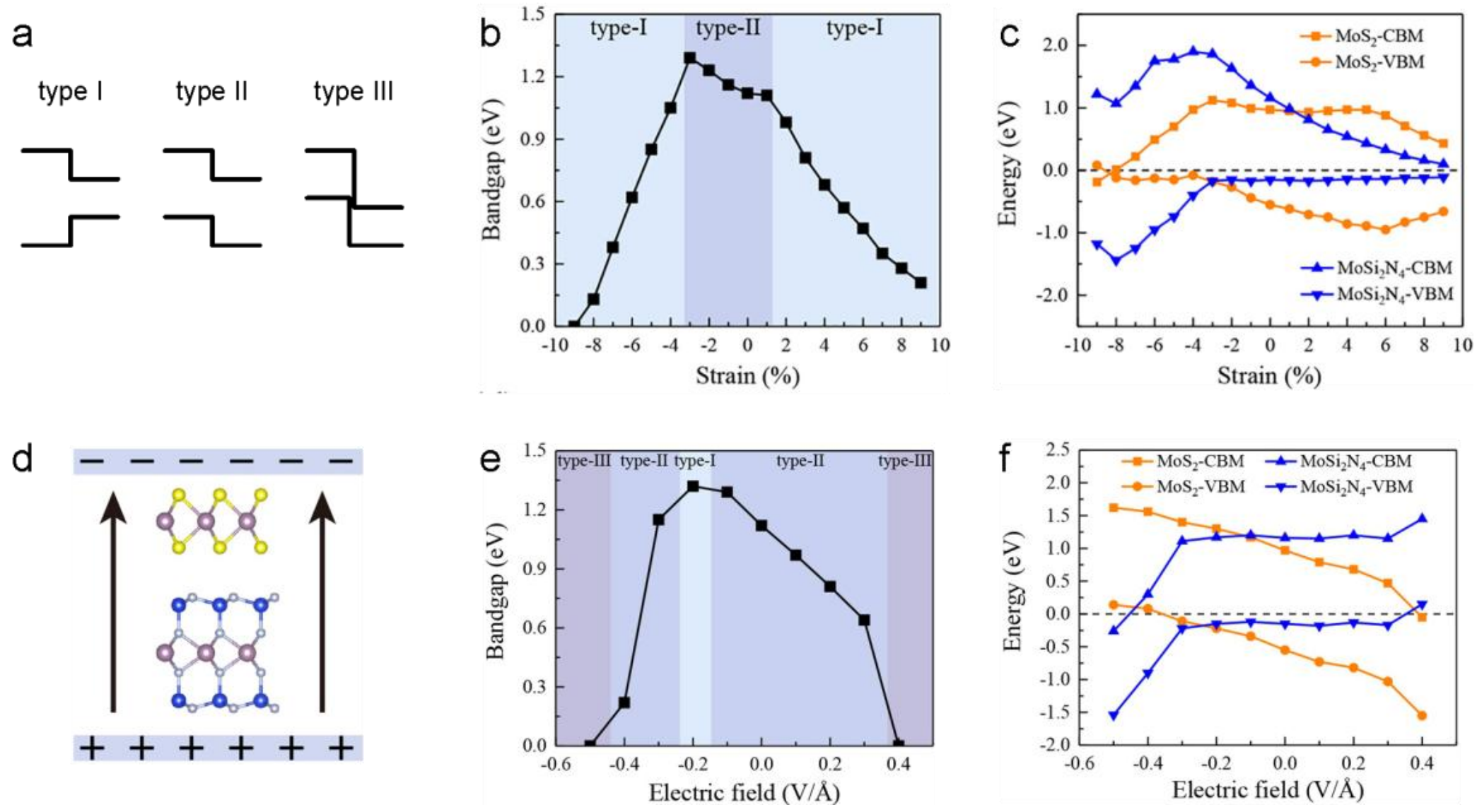

**Fig. 6 Band alignment and energy gaps in $MoSi_2N_4/MoS_2$ van der Waals heterostructure. a,** Three types of band alignment. **b – c,** Band gap and alignment type under biaxial strain. **d – f**, Band gap and alignment type under external electric field, (**d**) shows the direction of electric field. The Fermi level in (**c**) and (**f**) is set to zero. b – f adapted with permission from ref. 105.

An integrated system from such a vast library of $MA_2Z_4$ materials in combination with other 2D materials (such as graphene or TMDs) in a van der Waals heterostructure (vdWH) arrangement can exhibit optical absorption beneficial for realizing photodetectors with extended operational bandwidth. Various 2D materials with varying optical properties and conductivity can be employed in high-quality, ultraclean vertical all-2D vdWH devices of sizeable interfacial areas for broadband operation[78,113,114] (Bafekry2021NewJChem, Binh2021JPhysChemLett, Cai2022Crystals). In particular, a type-II band alignment between two adjacent semiconductors is desired for rapid separation of charge carriers before recombination across a heterointerface for photovoltaic applications. TMDs/$MA_2Z_4$ heterojunctions usually exhibit a type-I alignment (the alignment types are shown in Fig. 6a). However, applying an electric field or external strain can shift the alignment to staggered ones (type-II)[106] (Cai2021JMaterChem). Boron phosphide/$MoSi_2P_4$ or boron phosphide/$MoGe_2N_4$ are some of the direct band gap vdWHs that inherently form type-II band alignment[109,115] (Guo2022JPhysChem, Nguyen2021JPhysChemLett). Likewise, interlayer coupling and faster charge separation across a blue-P/$MoSi_2N_4$ type-II heterojunction are found to be beneficial for photocatalytic and photovoltaic applications[114] (Cai2022Crystals). $MoSi_2N_4/Cs_3Bi_2I_9$ heterostructure is reported to exhibit broadband absorption, which is absent in the individual materials[101] (Liu2022JAP). Hence, this class of materials is promising for integrated multicolor photonic devices such as photodetectors, photovoltaic cells, and optical modulators.

## 13 Catalytic properties

Many 2D materials have already proven themselves as excellent catalysts[116-119] (Deng2016NatNanotech, Deng2017NatCommun, Voiry2016AdvMat, Zhang2020ApplSurfSci). Due to the presence of transition metals[120-124] (Twilton2017NatRevChem, Peng2019JMatChemA, Yu2020AdvFunctMat, Zhang2021JMatChemA, Wang2022ApplSurfSci), $MA_2Z_4$ (M=Mo, W) materials can also be added to the line of the promising 2D candidates to serve as catalysts in water splitting, $CO_2$ reduction and other photochemical reaction due to their broadly tunable photochemical properties. In a photochemical reaction, a photon of light, absorbed by the catalyst, generates an electron-hole pair, these charges then migrate to the catalyst's surface where they participate in the chemical reaction. A suitable photocatalyst thus should: exhibit a large electrochemical active surface area (ECSA) – this requirement is fulfilled for all 2D materials; be stable at room temperature; its energy gap should be smaller than the energy of visible light photon (3.26 – 1.77 eV corresponding to 380 – 700 nm) in order to be able to absorb the light; its absorption coefficient in the visible light range should be comparable or exceeding $10^5$ $cm^{-1}$. In addition, the photocatalyst should have a built-in option for separating the created electrons and holes to avoid their recombination, and exhibit high charge carriers mobility. As suitable catalysts, semiconductors with direct band gap are preferred since they allow direct electron transition from valence to conductance band, without a need of phonon creation to fulfil the energy and momentum conservation laws. The bands edges of the catalyst should straddle the potentials of the reactions: its valence band edge (VBE) or valence band maximum (VBM) should be lower than the potential of an oxidation reaction, and its conductance band edge (CBE) or conduction band minimum (CBM) should be larger than the potential of a redox reaction, as illustrated in Fig. 7a. Reaction potential depends on pH as defined by the Nernst equation: $E^o(\text{pH}) = E^o(\text{pH=0}) + \text{pH}\cdot 0.059$ eV, (at 25° C). The band edges of a semiconductor can be evaluated either directly from its energy bands distribution, or from its bandgap $E_g$ alone by using the following equations[125] (Liu2011APL): $E_{\text{VBE}} = x - 4.5\text{eV} + 0.5E_g$, and $E_{\text{CBE}} = E_{\text{VBE}} - E_g$, where $x$ is the geometric average value of the electronegativity of the constituent atoms ($x$ = 5.195 for $MoSi_2N_4$[126] (Bartolotti1980JACS)), and 4.5 eV is the energy of free electrons on the hydrogen scale.

The water splitting reaction $H_2O \rightarrow H_2 + 1/2O_2$ consists of two half-reactions[127] (Nishioka2023NatRevMethods): hydrogen evolution reaction (HER) $2H^+ + 2e^- \rightarrow H_2$ and oxygen evolution reaction (OER) $H_2O + 2h^+ \rightarrow 1/2O_2 + 2H^+$. Thus, the catalyst's VBE should be lower than the potential of the OER ($E(H_2O/O_2)$ =-5.67 eV + pH·0.059 eV), its CBE should be higher than the potential of HER ($E(H^+/H_2)$ =-4.44 eV + pH·0.059 eV), with the minimal energy gap of 1.23 eV. Photocatalytic $CO_2$ reduction reaction can be realized by one of the following reactions:

$$CO_2 + 2H^+ + 2e \rightarrow HCOOH \qquad (E°=-0.194\ V)$$

$$CO_2 + 2H^+ + 2e \rightarrow CO + H_2O \qquad (E°=-0.099\ V)$$

$$CO_2 + 6H^+ + 6e \rightarrow CH_3OH + 2H_2O \qquad (E°=0.031\ V)$$

$$CO_2 + 8H^+ + 8e \rightarrow CH_4 + 2H_2O \qquad (E°=0.171\ V)$$

The potentials *E*° shown here (vs standard hydrogen electrode (SHE) at pH=0) were calculated using the free energy[128] (Latimer_book). The positions of the potentials for water splitting and $CO_2$ reduction reactions, together with the band edges of a hypothetically suitable catalyst are shown in Figure 7a.

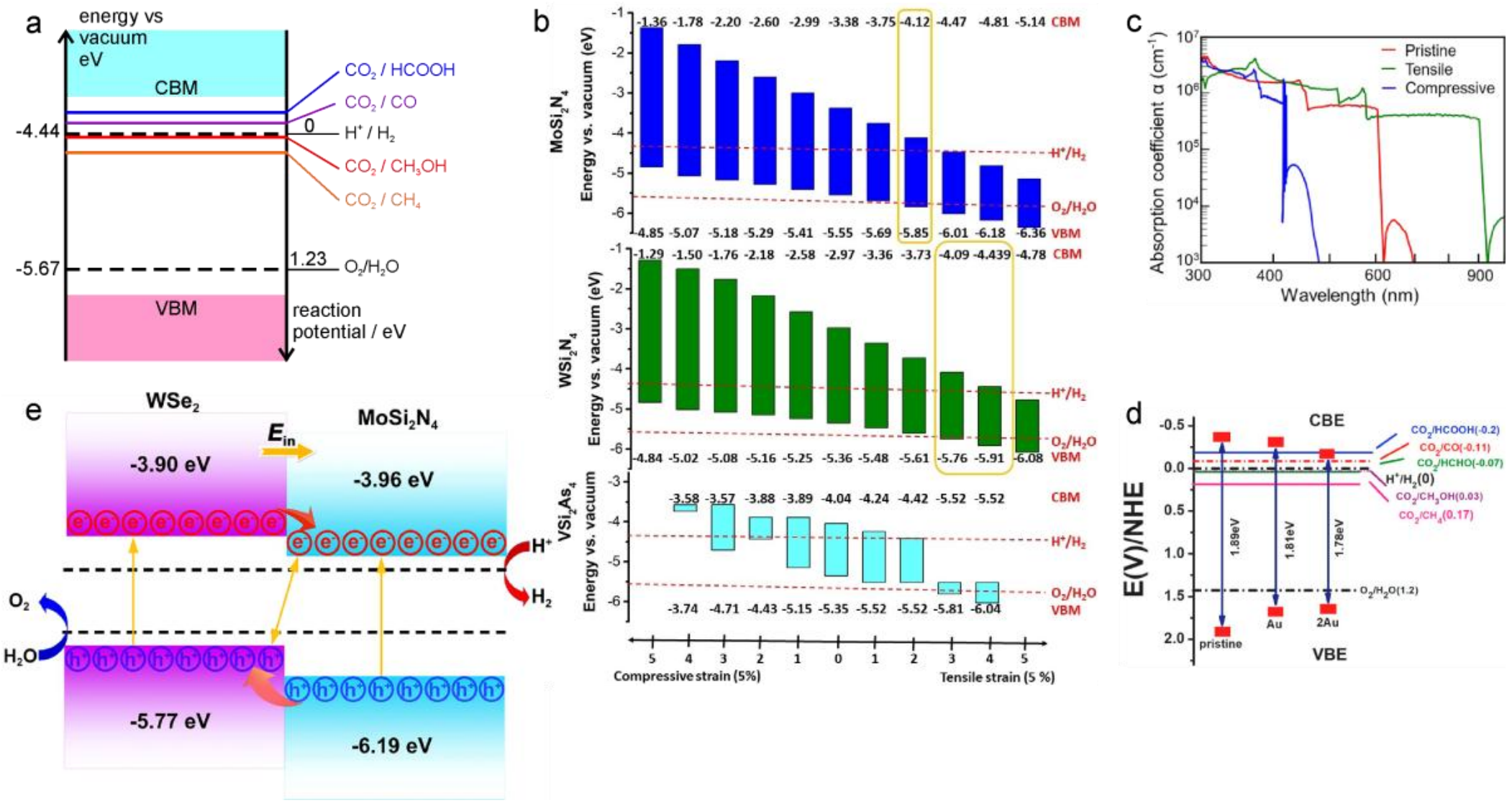

**Fig. 7 Photocatalytic properties of $MA_2Z_4$ materials. a,** Potentials of the water splitting and $CO_2$ reduction reactions, shown together with the conduction and valence band edges (CBE and VBE, respectively) of a hypothetical suitable catalyst. **b**, Calculated band edges of $WSi_2N_4$, $MoSi_2N_4$ and $VSi_2As_4$ at applied biaxial strain shown relatively to the vacuum level; adapted with permission from ref. [48]. **c**, Optical absorption coefficients of pristine and strained (ε = +6% and -6%) $MoSi_2N_4$ monolayers as a function of wavelength; adapted with permission from ref. 129. **d**, Band alignments of intrinsic $MoSi_2N_4$ and $MoSi_2N_4$-nAu (n =1, 2) for photocatalytic water splitting and carbon dioxide reduction. The band edges are given with respect to the NHE (normal hydrogen electrode) potential (V); adapted with permission from ref. [130]. **e**, Schematic diagram of the photocatalytic reaction for the $WSe_2/MoSi_2N_4$ type-II heterostructure. $E_{in}$ denotes the direction of the built-in electric field; adapted with permission from ref. [131].

The bandstructure of $MA_2Z_4$ materials suggest that some of these materials can be promising catalysts for photocatalytic water splitting or $CO_2$ reduction reactions. Moreover, biaxial strain can boost the photocatalytic properties by shifting the band edges[29,48] (Yu2021ApplMatInterfaces, Chen2021ChemistryAEuropeanJ) and increasing the absorption coefficient[129,132] (Jian2021JPhysChem, Xuefeng2022JPhysD). Chen et al investigated 54 $MA_2Z_4$ materials and found that only three materials have bandgaps larger than $E_g$ = 1.23 eV and thus can be suitable catalyst for water splitting reaction: $MoSi_2N_4$ ($E_g$ = 2.13 eV), $WSi_2N_4$ ($E_g$ =2.38 eV) and $VSi_2As_4$ ($E_g$ = 1.31 eV).

However, the band edges of these structures straddle the reaction potentials only when biaxial tensile strain is applied: $\varepsilon$ = 2% for $MoSi_2N_4$ and $\varepsilon$ = 3– 4% for $WSi_2N_4$)[48] (Chen2021ChemistryAEuropeanJ), Fig. 7b. The measured absorption coefficient of pristine $MoS_2N_4$ is close to zero at wavelengths exceeding 650 nm[13] (Hong2020Science), but can be increased by shifting the spectrum towards longer wavelengths when biaxial tensile strain of $\varepsilon$ = 6% is applied[129] (Jian2021JPhysChem), Fig. 7c.

In single-atom catalysts (SACs), transition metals (TM) are the most popular choice for SACs structures in $MA_2Z_4$ family, for their partially filled d-orbitals that allow for both donating and accepting electrons[133-135] (Sun2022PhysChemCh, Lu2022ApplSurfSci, Xun2023ApplEnMat). DFT study of 15 single metal atom-doped 2D $MA_2Z_4$ ($TM1/MA_2Z_4$) showed that $Cr_1/HfSi_2N_4$ is the ideal catalyst candidate for the NO reduction reaction (NORR) (Sun2022PhysChemCh). Using first-principle calculations, 5 (Sc, Ti, Fe, Co, Ni) out of 10 TMs on $MoSi_2N_4$ were identifies as suitable SACs for $CO_2$ reduction[135] (Xun2023ApplEnMat). Alkali-metal (Li, Na, or K) adsorbed on $MoSi_2N_4$ greatly increase the absorption coefficient $\alpha$ in the visible light range (380 – 780nm) when compared to pristine $MoSi_2N_4$: $\alpha$ = 0.34 – 0.02 (pristine), 0.57 – 0.92 (Li), 1.09 – 1.34 (Na), and 0.89 – 1.14 (K) $\times 10^5$ $cm^{-1}$ (Fig. 7d), respectively[136] (Sun2022CommunTheorPhys). Similarly, adding one or two Au atoms to $MoSi_2N_4$ increases $\alpha$ (Fig. 7e), however adding more than 3 Au atoms decreases the band gap, eventually turning it to zero when 9 Au atoms are added[130] (Xu2022JPhysChemSolids).

In pristine $MoSi_2N_2$, both VBM and CBM are mainly contributed by Mo atoms[130] (Xu2022JPhysChemSolids), thus creating a high probability for recombination of the photo-generated electrons and holes. This issue can be resolved by placing $MoSi_2N_2$ into a van der Waals structure, where the redistributed charge creates a built-in electric field that can effectively separate the photo-generated charges. In van der Waals structures, the atoms that contribute to VBM and CBM can be from different, spatially separated layers, which leads to a better separation between photo-generated charges. For example, in AB-stacking bilayer $MoSi_2N_4$, the CBM and VBM are dominated by the states of the Mo atoms in the lower and upper layer, respectively[137] (Zhao2021JPHysChemLett). The built-in electric field transports the charges to the catalyst surface, and two parts of a reaction (oxidation and reduction) can run on the two sides of the catalyst (Fig. 6f). Besides bilayer $MoSi_2N_4$ and $(MoSi_2N4)_{5-n}/(MoSiGeN_4)_n$ (n=0,...5)[138] (Mwankemwa2022ResPhys), the following $MoS_2N_4$-based vdWH were proposed as efficient photocatalysts: $C_2N/MoS_2N_4$ for HER reaction[139] (Zeng2021PhysChemChemPhys), $MoS_2N_4/CrS_2$[140] (Li2022PhysicaE), $InSe/MoS_2N_4$[141] (He2022PhysChemChemPhys), $Mo_S2N_4$/BlueP[132] (Xuefeng2022JPhysDApplPhys) and $WS_2/MoS_2N_4$[131] (Liu2023ApplSurfSci) for photocatalytic water splitting. The catalytic properties of $MA_2Z_4$ materials are summarized in Table S3.

In Janus structures $MSiGeN_4$ (M = Mo and W)[28] (Guo2021JMaterChemC), breaking the vertical symmetry of the structure creates a significant built-in electric field which can enhance the charge separation. Janus structures $MoSiGeN_4$ and $WSiGeN_4$ were found as potential catalysts for photocatalytic water splitting[29] (Yu2021ApplMatInterf). Zhang et al screened 104 Janus-$MA_2Z_4$ structures and identified 13 thermally and environmentally stable structures; from which H-$HfSiGeP_4$, H$MoSiGeP_4$, T-$ScSiGeN_4$, and T-$ZrSiGeN_4$ are suitable photocatalysts for $CO_2$ reduction reactions[142] (Zhang2022FrontPhys). The photocatalytic properties of $MA_2Z_4$ materials are summarized in Table S1. In addition, DFT studies by Chen et al identified 42 dynamically stable $MA_2Z_4$ (M = Ti, Zr, Hf, V, Nb, Ta, Cr, Mo, W; A = Si or Ge; Z = N, P, or As) monolayers, where four of them, $VGe_2As_4$, $CrGe_2As_4$, $VSi_2As_4$, and $NbSi_2As_4$, are predicted to exhibit promising oxygen reduction reaction (ORR) electrocatalytic activity[48] (Chen2021ChemA).

Relatively small negative adsorption energies of environmental gas molecules ($H_2$, $N_2$, NO, CO, $O_2$, CO2, $NO_2$, $SO_2$, $H_2O$, $H_2S$, $NH_3$, and $CH_4$) on $MoSi_2N_4$ surface (physisorption) suggest the possibility of using $MoSi_2N_4$ monolayers as gas sensors[143,144] (Xiao2022ACS_Omega, Bafekry2021ApplSurfSci). The abovementioned optical properties of $MoSi_2N_4$ and their heterostructures – appropriate band gap, high absorption coefficient, high carriers' mobility – make these materials promising candidates for solar cells. The predicted photoelectric conversion efficiency (PCE) for the BP/$MoSi_2P_4$ heterostructure of 22.2% is the largest PCE among 2D heterostructures[115] (Guo2022JPhysChemC).

# 14 Conclusions

The new class of septuple layer two-dimensional material will offer new opportunities and functionalities when searching for new quantum phenomena in transport and optical properties of such materials, as well as investigating new van der Waals heterostructures. Crystals with a range of properties (metallic, semiconducting, ferromagnetic, superconducting, etc) have been predicted to be the members of the class. What is unique about such materials, is that the seven atomic layers thick unit cell offers a lot of variation both in terms of chemical composition and the crystal phase formation, thus allowing for a lot of tunability as well as observation of multiple phase transitions. We envisage that the multitude of the phases available can result in structural and electronic phase transitions observed under ambient conditions, which might be important for novel electronic (like neuromorphic computing, for instance). Thus, in optical applications, $MA_2Z_4$ can act as a polarization-sensitive element with broadband responsivity spanning across visible to mid-IR frequency domains while some semiconducting $MA_2Z_4$ materials with magnetic order can offer novel spintronic applications. Furthermore, the relatively large thickness of the crystals can allow the generation of transversal electric field, which can non-trivially alter the electronic, optical, mechanical and other properties. This is especially important for photo-catalysis, since it allows for efficient electron-hole separation, which is not observed in other 2D materials. The first member of the group - $MoSi_2N_4$ – was recently synthesised using seemingly simple technique. Still, despite the plethora of intriguing predictions – the experimental growth of other members of the family is lugging behind. At the same time, there is no indication that there are any fundamental limitations which can impede the progress in this area, so we can expect more similar crystals to appear in the near future. As this family of materials has a very rich compositional and structural variety, the use of machine learning for the design and synthesis of such crystals might be beneficial, as indeed has already been implemented in the other areas of material science[145,146] (Merchant2023Nature, Hippalgoankar2023NatRevMat), including 2D materials[147-149] (Kazeev2023npjCompMat, Huang2023npj2DMatAppl, Sun2023ACSApplMatInterf). We hope that this review will instigate further experimental activity in this area.

Supplementary Information for

# $MoSi_2N_4$-like crystals – the new family of two-dimensional materials

T. Latychevskaia[1], D. Bandurin[2]*, K. S. Novoselov[3]*

*[1]Paul Scherrer Institute, Forschungsstrasse 111, 5232 Villigen, Switzerland*
*[2]Department of Materials Science and Engineering, National University of Singapore, Singapore, 117575, Singapore*
*[3]Institute for Functional Intelligent Materials, National University of Singapore, Building S9, 4 Science Drive 2, Singapore 117544*
**E-mail: dab@nus.edu.sg; kostya@nus.edu.sg*

## Contents

## 1. $MA_2Z_4$ structures

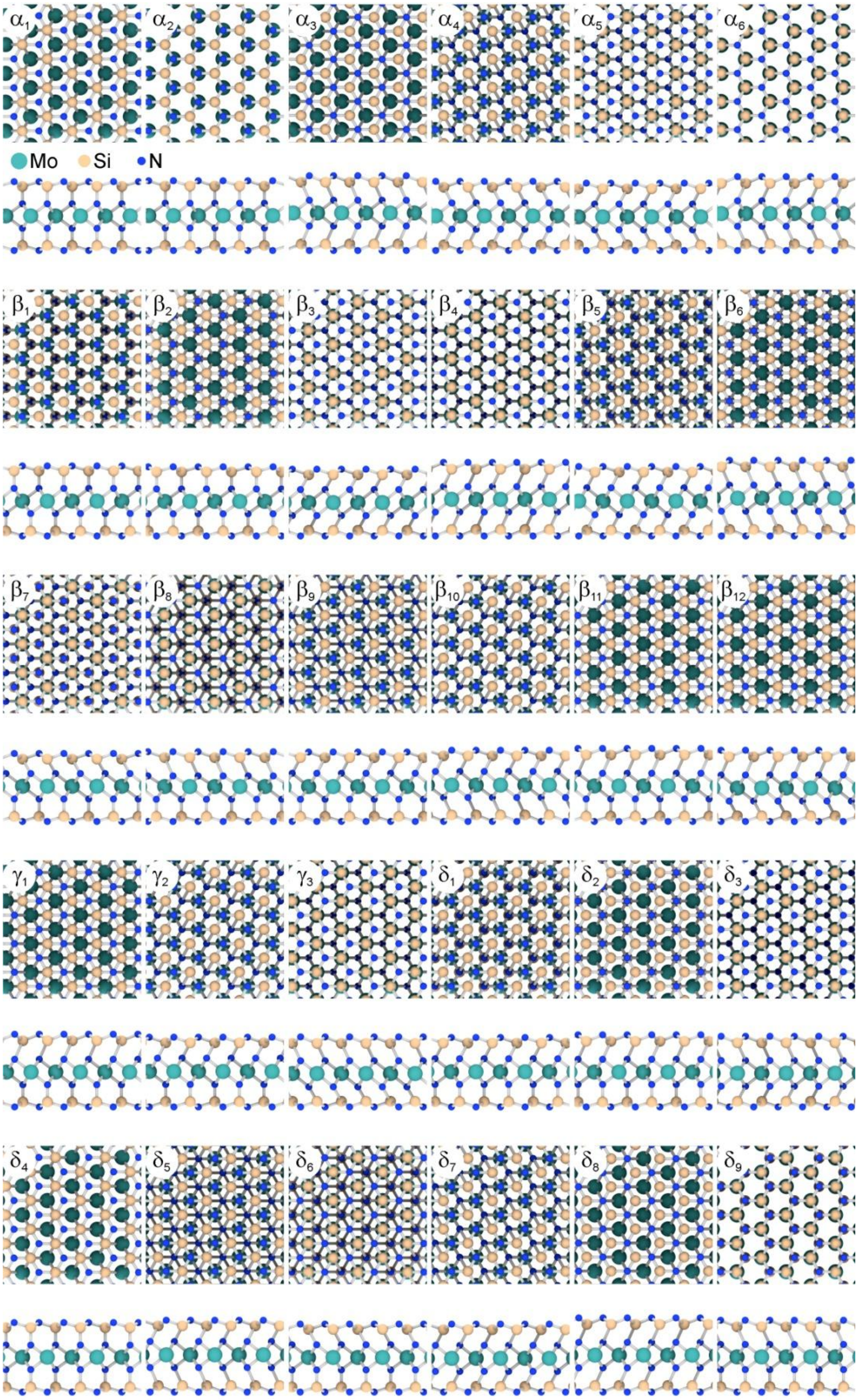

**Fig. S1**. 30 possible structures of $MA_2Z_4$ calculated for $MoSi_2N_4$, using atomic coordinates provided in ref. [1] (Wang2021NatCommun); top and side views for each structure.

## 2. $MA_2Z_4$-metal structures

### 2.1 Tunnelling probability

The tunnelling probability of a particle with energy *E* through a potential barrier *V*(*x*) can be calculated as:

$$T(E)=\exp\left\{-2\int_{x_1}^{x_2}\sqrt{\frac{2m_e}{\hbar^2}\left[V(x)-E\right]}\mathrm{d}x\right\}\approx\exp\left[-2w\sqrt{\frac{2m_e}{\hbar^2}(V_0-E)}\right], \quad \text{(S1)}$$

where a rectangular barrier of height $V_0$ and width *w* is assumed, $m_e$ is the free-electron mass and $\hbar$ is the reduced Planck's constant. The tunnelling probability through the Schottky barrier is given by

$$P_{\mathrm{TB}}(w_{\mathrm{TB}},\Phi_{\mathrm{TB}})=\exp\left(-2w_{\mathrm{TB}}\frac{\sqrt{2m_e\Phi_{\mathrm{TB}}}}{\hbar}\right)=\exp(-2w_{\mathrm{TB}}k_{\mathrm{TB}}), \quad \text{(S2)}$$

where $\Phi_{\mathrm{TB}}$ and $w_{\mathrm{TB}}$ are the height and width of the barrier, $k_{\mathrm{TB}}=\sqrt{2m_e\Phi_{\mathrm{TB}}}/\hbar$ is the wavenumber.

### 2.2 Specific resistivity

The relationship between the current density and voltage in a tunnel junction is given by Eq. (20) in ref.[2] (Simmons1963JAP_p1793)

$$J_{\mathrm{TB}}=\frac{e}{2\pi h w_{\mathrm{TB}}^2}\left\{\bar{\varphi}\exp\left(-A\bar{\varphi}^{1/2}\right)-(\bar{\varphi}+eV)\exp\left[-A(\bar{\varphi}+eV)^{1/2}\right]\right\}, \quad \text{(S3)}$$

where $\bar{\varphi}$ is the mean value of the barrier height above the Fermi level of the negatively biased electrode, *V* is the applied voltage, *e* is the electron charge, and

$$A=\frac{4\pi w_{\mathrm{TB}}}{h}(2m_e)^{1/2}=\frac{2w_{\mathrm{TB}}}{\hbar}(2m_e)^{1/2}. \quad \text{(S4)}$$

At the intermediate voltages, the mean barrier height is given by

$$\bar{\varphi}=\frac{\varphi_1+\varphi_2-eV}{2}, \quad \text{(S5)}$$

and for $\varphi_1=\varphi_2=\Phi_{\mathrm{TB}}$, Eq. (S3) becomes:

$$J_{\mathrm{TB}}=\frac{e}{4\pi^2\hbar w_{\mathrm{TB}}^2}\times$$

$$\left\{\left(\Phi_{\mathrm{TB}}-\frac{eV}{2}\right)\exp\left(-2w_{\mathrm{TB}}\frac{\sqrt{2m_e(\Phi_{\mathrm{TB}}-eV/2)}}{\hbar}\right)-\left(\Phi_{\mathrm{TB}}+\frac{eV}{2}\right)\exp\left[-2w_{\mathrm{TB}}\frac{\sqrt{2m_e(\Phi_{\mathrm{TB}}+eV/2)}}{\hbar}\right]\right\} \quad \text{(S6)}$$

The tunnelling resistivity is calculated as:

$$\rho_{\mathrm{TB}}(w_{\mathrm{TB}},\Phi_{\mathrm{TB}})=\left(\frac{\mathrm{d}J_{\mathrm{TB}}}{\mathrm{d}V}\right)^{-1}\Bigg|_{V=0}=\frac{4\pi^2\hbar w_{\mathrm{TB}}^2}{e^2}\frac{\exp(2w_{\mathrm{TB}}k_{\mathrm{TB}})}{w_{\mathrm{TB}}k_{\mathrm{TB}}-1}. \quad \text{(S7)}$$

For thin barriers, Eq. (S7) gives unphysical negative values, and instead, a low-voltage ($eV\simeq 0$) approximation is used. For low voltages, Eq. (S3) reduces to

$$J_{\mathrm{TB}} = \frac{e^2\left(2m_e\bar{\varphi}\right)^{1/2}}{4\pi^2\hbar^2 w_{\mathrm{TB}}} V \exp\left(-2w_{\mathrm{TB}}\frac{\left(2m_e\bar{\varphi}\right)^{1/2}}{\hbar}\right), \quad \text{(S8)}$$

and substituting

$$\bar{\varphi} = \frac{\varphi_1+\varphi_2}{2} = \Phi_{\mathrm{TB}} \quad \text{(S9)}$$

gives

$$J_{\mathrm{TB}} = \frac{e^2\left(2m_e\Phi_{\mathrm{TB}}\right)^{1/2}}{4\pi^2\hbar^2 w_{\mathrm{TB}}} V \exp\left(-2w_{\mathrm{TB}}\frac{\left(2m_e\Phi_{\mathrm{TB}}\right)^{1/2}}{\hbar}\right), \quad \text{(S10)}$$

The tunnelling resistivity is calculated as:

$$\rho_{\mathrm{TB}}\left(w_{\mathrm{TB}},\Phi_{\mathrm{TB}}\right) = \left(\frac{dJ_{\mathrm{TB}}}{dV}\right)^{-1}\Bigg|_{V=0} = \frac{4\pi^2\hbar^2 w_{\mathrm{TB}}}{e^2\left(2m_e\Phi_{\mathrm{TB}}\right)^{1/2}} \exp\left(2w_{\mathrm{TB}}\frac{\left(2m_e\Phi_{\mathrm{TB}}\right)^{1/2}}{\hbar}\right), \quad \text{(S11)}$$

note that the result in Eq. (S11) is different from the result obtained by Simmons by factor 2/3, the issue has been previously discussed in ref.[3] (Matthews2018JAP).

## 2.3 Summary of electronic properties of $MA_2Z_4$-metal structures

**Table S1**. Properties of $MA_2Z_4$-metal systems. "G"=graphene. n-type and p-type are highlighted in blue and pink, Schottky and Ohmic - in weak and strong color, respectively. The tunnelling barrier probability and resistivity were calculated in this study by using Eqs. (S2) and (S7), respectively. When the resistivity by Eq. (S7) was negative, it was re-calculated by Eq. (S11) (highlighted in lilac). LM - lattice mismatch (copied from the cited papers), $d_0$ - interlayer distance, EF - the electric field.

| Composition | Schottky barrier height | | tunneling barrier | | | Resistivity $10^{-9}$ Ohm·cm$^2$ | LM % | $d_0$/ Å | strain | | EF | reference |
|---|---|---|---|---|---|---|---|---|---|---|---|---|
| | $\Phi_{Bn}$ eV | $\Phi_{Bp}$ eV | height eV | width Å | probability % | | | | vertical | biaxial | | |
| **$MoSi_2N_4$** | | | | | | | | | | | | |
| G | 1.027 | 0.979 | NA | NA | NA | NA | 2.39 | 3.35 | NA | p to n at ε<-2%, to Ohmic at ε<-10% | NA | [4] |
| G | 0.922 | 0.797 | NA | NA | NA | NA | 1.6 | 3.302 | p to n at $d$>3.5Å | NA | p to n at EF>0.05V/Å | [5] |
| G | 1.49 | 0.96 | NA | NA | NA | NA | 1.65 | 3.16 | NA | NA | NA | [6] |
| G | 1.03 | 0.98 | 4.04 | 1.92 | 1.917 | 3.191 | 2.39 | 3.35 | NA | NA | NA | [7] |
| G | 0.68 | 0.97 | 9.98 | 1.30 | 1.488 | 1.669 | 1.59 | 3.18 | n to p at $d$<2.89Å | NA | n to p at EF>0.06 V/Å, to Ohmic at EF>0.8 V/Å and EF<-0.6 V/Å | [8] |
| G | 1.14 | 1.06 | NA | NA | NA | NA | 1.70 | 3.36 | p to n at $d$>3.4Å | NA | p to n at EF>0, Ohmic at EF>0.3V/Å and EF<-0.3V/Å | [9] |
| G | 1.28 | 0.38 | 4.45 | 1.96 | 1.446 | 3.854 | 1.60 | 3.33 | NA | NA | NA | [10] |
| G | NA | 0.95 | NA | NA | NA | NA | 1.07 | 3.30 | NA | NA | NA | [11] |
| $Hf_4C_3$ | -0.915 | 2.711 | 0.000 | 0.000 | 100 | 0.000 | 1.5 | 2.25 | NA | NA | NA | [12] |
| $Hf_4C_3F_2$ | -0.118 | 1.921 | 2.304 | 1.283 | 13.596 | 0.197 | 1.8 | 2.55 | NA | NA | NA | [12] |
| $Hf_4C_3O_2$ | 1.500 | 0.543 | 3.981 | 1.493 | 4.725 | 1.454 | 1.2 | 2.65 | NA | NA | NA | [12] |
| $Hf_4C_3O_2H_2$ | 0.543 | 2.174 | 0.419 | 0.487 | 72.397 | 0.033 | 1.2 | 1.80 | NA | NA | NA | [12] |
| 1T-$MoS_2$ | NA | 0.76 | NA | NA | NA | NA | 2.61 | 3.20 | NA | NA | NA | [11] |
| MoSH | 0.54 | 0.50 | NA | NA | NA | NA | 1.70 | 2.90 | p to n at Δ$d$<-0.1Å, to Ohmic at Δ$d$<-1.2Å | NA | p to n at EF>0 V/nm, to Ohmic at EF<-0.35 V/nm | [13] |
| MoHS | 0.90 | 0.11 | NA | NA | NA | NA | 1.70 | 3.32 | NA | NA | p to n at EF>0.3 V/nm, to Ohmic at EF<-0.3 V/nm | [13] |
| $NbS_2$ | NA | -0.08 | NA | NA | NA | NA | 0.19 | 3.10 | NA | NA | NA | [11] |
| $NbS_2$ | 1.642 | 0.042 | NA | NA | NA | NA | 0.07 | 3.192 | remains p | NA | p to Ohmic at EF<-0.2 V/Å | [5] |
| $NbS_2$ | 1.87 | -0.04 | 5.21 | 1.93 | 1.096 | 4.387 | 0.85 | 3.18 | NA | NA | NA | [7] |
| $NbS_2$ | 1.65 | -0.03 | 5.14 | 1.88 | 1.269 | 3.818 | 0.14 | 3.15 | NA | NA | NA | [10] |
| $NbSe_2$ | NA | -0.01 | NA | NA | NA | NA | 1.77 | 3.20 | NA | NA | NA | [11] |
| $NbSe_2$ | 1.63 | 0.09 | 4.74 | 1.87 | 1.543 | 3.386 | 3.69 | 3.26 | NA | NA | NA | [7] |
| $Nb_4C_3$ | -0.109 | 2.418 | 0.000 | 0.000 | 100 | 0.000 | 4.5 | 2.75 | NA | NA | NA | [12] |
| $Nb_4C_3F_2$ | 1.217 | 0.970 | 3.901 | 1.733 | 2.998 | 2.156 | 3.7 | 2.80 | NA | NA | NA | [12] |
| $Nb_4C_3O_2$ | 2.217 | 0.075 | 4.723 | 1.768 | 1.951 | 2.683 | 4 | 2.85 | NA | NA | NA | [12] |
| $Nb_4C_3O_2H_2$ | -0.065 | 2.489 | 0.902 | 0.856 | 43.474 | 0.066 | 3.7 | 2.00 | NA | NA | NA | [12] |
| $TaS_2$ | 1.86 | -0.03 | 5.12 | 1.91 | 1.193 | 4.083 | 0.45 | 3.18 | NA | NA | NA | [7] |
| $TaSe_2$ | 1.86 | -0.03 | 5.14 | 1.93 | 1.129 | 3.187 | 3.27 | 3.20 | NA | NA | NA | [7] |
| $Ti_2NF_2$ | 1.40 | 0.31 | 5.48 | 1.56 | 2.371 | 1.911 | 0.52 | 2.61 | NA | NA | NA | [14] |
| $Ti4C_3$ | -0.396 | 1.393 | 0.124 | 0.245 | 91.540 | 0.024 | 4 | 2.30 | NA | NA | NA | [12] |
| $Ti_4C_3F_2$ | 0.639 | 0.473 | 4.057 | 2.037 | 1.494 | 4.089 | 2.8 | 3.10 | NA | NA | NA | [12] |
| $Ti_4C_3O_2$ | 1.791 | -0.170 | 5.002 | 1.967 | 1.102 | 4.540 | 3.7 | 3.15 | NA | NA | NA | [12] |
| $Ti_4C_3O_2H_2$ | -0.446 | 1.483 | 0.684 | 0.603 | 59.990 | 0.038 | 3.9 | 1.90 | NA | NA | NA | [12] |
| $Ti_4N_3O_2$ | -0.26 | 2.11 | 4.72 | 1.61 | 2.777 | 1.912 | 2.65 | 2.65 | NA | NA | NA | [14] |
| $Ti_4N_3(OH)_2$ | -0.26 | 2.11 | 0 | 0 | 100 | 0.000 | 3.99 | 1.89 | NA | NA | NA | [14] |
| $VS_2$ | 1.96 | -0.02 | 5.29 | 1.95 | 1.010 | 4.706 | 3.38 | 3.23 | NA | NA | NA | [7] |
| $VSe_2$ | 1.94 | 0.53 | 4.40 | 1.88 | 1.759 | 3.195 | 1.89 | 3.29 | NA | NA | NA | [7] |
| $V_2CF_2$ | 1.83 | 0.02 | 5.04 | 1.61 | 2.464 | 2.003 | 3.44 | 2.59 | NA | NA | NA | [14] |
| $V_2C(OH)_2$ | -0.20 | 2.06 | 0 | 0 | 100 | 0.000 | 3.44 | 1.92 | NA | NA | NA | [14] |
| $V_3C_2O_2$ | 1.92 | -0.20 | 4.12 | 1.53 | 4.150 | 1.548 | 4.74 | 2.64 | NA | NA | NA | [14] |
| $V_4C_3$ | -0.151 | 1.779 | 0.106 | 0.182 | 94.109 | 0.019 | 0.2 | 2.20 | NA | NA | NA | [12] |
| $V_4C_3F_2$ | 1.445 | 0.100 | 5.015 | 2.044 | 0.919 | 5.484 | 0.1 | 3.00 | NA | NA | NA | [12] |
| $V_4C_3O_2$ | 1.657 | -0.085 | 5.573 | 2.023 | 0.750 | 6.121 | 0.8 | 3.05 | NA | NA | NA | [12] |
| $V_4C_3O_2$ | 1.79 | -0.10 | 5.18 | 1.65 | 2.133 | 2.241 | 4.78 | 2.67 | NA | NA | NA | [14] |
| $V_4C_3O_2H_2$ | -0.375 | 1.913 | 0.637 | 0.680 | 57.345 | 0.047 | 0.9 | 1.95 | NA | NA | NA | [12] |
| 1T-$WS_2$ | NA | 0.55 | NA | NA | NA | NA | 3 | 3.10 | NA | NA | NA | [11] |
| $W_2NF_2$ | 1.83 | 0.02 | 5.34 | 1.55 | 2.548 | 1.832 | 0.52 | 2.64 | NA | NA | NA | [14] |
| $W_3N_2(OH)_2$ | -0.12 | 1.99 | 0 | 0 | 100 | 0.000 | 1.34 | 1.79 | NA | NA | NA | [14] |
| $Zr_4C_3$ | 1.913 | 2.366 | 0.000 | 0.000 | 100 | 0.000 | 0.3 | 2.40 | NA | NA | NA | [12] |
| $Zr_4C_3F_2$ | -0.164 | 1.879 | 2.866 | 1.626 | 5.958 | 1.754 | 0.2 | 2.80 | NA | NA | NA | [12] |
| $Zr_4C_3O_2$ | 1.753 | 0.098 | 4.511 | 1.847 | 1.796 | 3.051 | 0.9 | 2.85 | NA | NA | NA | [12] |
| $Zr_4C_3O_2H_2$ | -0.561 | 2.281 | 0.716 | 0.749 | 52.236 | 0.054 | 0.2 | 1.95 | NA | NA | NA | [12] |
| **$WSi_2N_4$** | | | | | | | | | | | | |
| G | 1.557 | 0.776 | NA | NA | NA | NA | 2.39 | 3.37 | NA | p to n at $\varepsilon$<-4%, to Ohmic at $\varepsilon$<-10% | NA | [4] |
| G | 1.56 | 0.78 | 4.04 | 1.89 | 2.040 | 3.002 | 2.39 | 3.37 | NA | NA | NA | [7] |
| G | 1.67 | 0.85 | NA | NA | NA | NA | 1.70 | 3.34 | remains p type | NA | p to n at EF>0.14V/Å, Ohmic at EF<-0.2V/A and >0.4V/Å | [15] |
| G | 1.79 | 0.22 | 4.39 | 1.96 | 1.488 | 3.793 | 1.67 | 3.32 | NA | NA | NA | [10] |
| G | NA | 0.74 | NA | NA | NA | NA | 1.12 | 3.30 | NA | NA | NA | [11] |
| 1T-$MoS_2$ | NA | 0.50 | NA | NA | NA | NA | 2.66 | 3.10 | NA | NA | NA | [11] |
| $NbS_2$ | NA | -0.08 | NA | NA | NA | NA | 0.24 | 3.10 | NA | NA | NA | [11] |
| $NbS_2$ | 2.00 | -0.04 | 5.04 | 1.84 | 1.452 | 3.389 | 0.21 | 3.15 | NA | NA | NA | [10] |
| $NbS_2$ | 2.18 | -0.06 | 5.12 | 1.92 | 1.166 | 4.183 | 0.85 | 3.18 | NA | NA | NA | [7] |
| $NbSe_2$ | NA | -0.02 | NA | NA | NA | NA | 1.72 | 3.20 | NA | NA | NA | [11] |
| $NbSe_2$ | 2.02 | -0.01 | 4.68 | 1.88 | 1.549 | 3.414 | 3.69 | 3.26 | NA | NA | NA | [7] |
| $TaS_2$ | 2.18 | -0.05 | 5.02 | 1.90 | 1.275 | 3.887 | 0.45 | 3.19 | NA | NA | NA | [7] |
| $TaSe_2$ | 1.98 | 0.02 | 4.60 | 1.87 | 1.642 | 3.276 | 3.27 | 3.28 | NA | NA | NA | [7] |
| $VS_2$ | 2.29 | -0.05 | 5.19 | 1.96 | 1.030 | 4.696 | 3.38 | 3.25 | NA | NA | NA | [7] |
| $VSe_2$ | 1.84 | 0.34 | 4.39 | 1.88 | 1.767 | 3.187 | 1.89 | 3.30 | NA | NA | NA | [7] |
| 1T-$WS_2$ | NA | 0.36 | NA | NA | NA | NA | 3.05 | 3.20 | NA | NA | NA | [11] |
| | | | | | | | | | | | | |
| G/$MoSi_2P_4$ | 0.13 | 0.60 | 6.11 | 1.04 | 7.179 | 0.771 | 5.35 | 3.33 | n to p at *d*<2.87Å, to Ohmic at *d*<2.3Å | NA | n to p at EF>0.28 V/Å, to Ohmic at EF<-0.4 V/Å and EF>1.0V/Å | [8] |
| G/$MoSi_2As_4$ | 0.32 | 0.31 | 5.67 | 1.31 | 4.092 | 1.137 | 2.54 | 3.51 | n to p at *d*<3.54Å, to ohmic at *d*<2.4Å | NA | p to n at EF<0, to Ohmic at EF<-1.6 V/Å and EF>1.0V/Å | [8] |
| $MoGe_2N_4$/G | 0.63 | 1.54 | NA | NA | NA | NA | 1.8 | 3.38 | n to p at *d*<2.8Å | to Ohmic at $\varepsilon$>6% | n to p at EF<-0.3V/Å, to Ohmic at EF>0.46V/Å | [6] |
| | | | | | | | | | | | | |
| G/MoGeSiN$_4$ | 0.63 | 1.13 | NA | NA | NA | NA | 1.20 | 3.27 | n to p at $\Delta d$<-0.45Å | NA | n to p at EF>0.2V/Å, to Ohmic at EF<-0.4V/Å | [16] |
| G/MoSiGeN$_4$ | 1.03 | 0.74 | NA | NA | NA | NA | 1.20 | 3.33 | p to n at $\Delta d$>0.35Å | NA | p to n at EF<-0.15V/A, Ohmic at EF>0.4V/Å | [16] |
| **$MoSi_2N_4$** | | | | | | | | | | | | |
| Ag | 0.12 | 1.55 | 3.04 | 1.48 | 7.107 | 1.552 | 0.49 | 3.12 | NA | NA | NA | [10] |
| Au | 1.19 | 0.59 | 3.73 | 1.55 | 4.655 | 1.569 | 0.61 | 3.13 | NA | NA | NA | [10] |
| Cu | 0.11 | 1.60 | 2.79 | 1.23 | 12.183 | 3.832 | 0.92 | 2.82 | NA | NA | NA | [10] |
| In | -0.04 | 1.70 | 2.97 | 1.69 | 5.058 | 1.861 | 1.53 | 3.40 | NA | NA | NA | [10] |
| Ni | 0.16 | 1.57 | 2.10 | 1.29 | 14.728 | 0.191 | 0.74 | 2.37 | NA | NA | NA | [10] |
| Sc | -0.24 | 1.93 | 0.00 | 0.00 | 100 | 0.000 | 1.04 | 2.31 | NA | NA | NA | [10] |
| Ti | -0.14 | 1.81 | 0.00 | 0.00 | 100 | 0.000 | 0.93 | 1.75 | NA | NA | NA | [10] |
| Pd | 0.70 | 0.99 | 3.09 | 1.13 | 13.064 | 8.984 | 3.65 | 2.73 | NA | NA | NA | [10] |
| Pt | 1.23 | 0.45 | 4.06 | 1.47 | 4.808 | 1.409 | 3.11 | 3.02 | NA | NA | NA | [10] |
| **$WSi_2N_4$** | | | | | | | | | | | | |
| Ag | 0.67 | 1.37 | 3.01 | 1.42 | 8.011 | 1.557 | 0.55 | 3.09 | NA | NA | NA | [10] |
| Au | 1.44 | 0.68 | 3.73 | 1.48 | 5.346 | 1.431 | 0.68 | 3.13 | NA | NA | NA | [10] |
| Cu | 0.65 | 1.48 | 2.80 | 1.23 | 12.137 | 3.713 | 0.88 | 2.82 | NA | NA | NA | [10] |
| In | 0.43 | 1.59 | 2.94 | 1.69 | 5.135 | 1.862 | 1.53 | 3.42 | NA | NA | NA | [10] |
| Ni | 0.82 | 1.13 | 2.08 | 1.29 | 14.863 | 0.191 | 0.82 | 2.38 | NA | NA | NA | [10] |
| Sc | -0.06 | 2.03 | 0.00 | 0.00 | 100 | 0.000 | 1.11 | 2.33 | NA | NA | NA | [10] |
| Ti | -0.12 | 2.06 | 0.00 | 0.00 | 100 | 0.000 | 0.85 | 1.78 | NA | NA | NA | [10] |
| Pd | 1.14 | 0.97 | 2.98 | 1.13 | 13.551 | 0.153 | 3.72 | 2.71 | NA | NA | NA | [10] |
| Pt | 1.85 | 0.20 | 4.04 | 1.46 | 4.945 | 1.389 | 3.18 | 3.02 | NA | NA | NA | [10] |
| **$CrSi_2N_4$** | | | | | | | | | | | | |
| G | 0.03 | 0.45 | 4.35 | 1.89 | 1.761 | 3.226 | 0.15 | 3.35 | NA | NA | NA | [17] |
| Ag | -0.03 | 0.54 | 2.91 | 1.30 | 10.308 | 1.953 | 0.04 | 2.79 | NA | NA | NA | [17] |
| Au | 0.02 | 0.54 | 3.69 | 1.44 | 5.876 | 1.372 | 0.19 | 3.02 | NA | NA | NA | [17] |
| Cu | -0.03 | 0.52 | 2.79 | 1.09 | 15.482 | 0.133 | 0.55 | 2.62 | NA | NA | NA | [17] |
| Ni | -0.01 | 0.60 | 3.17 | 1.20 | 11.201 | 2.204 | 0.62 | 2.67 | NA | NA | NA | [17] |
| Pd | -0.02 | 0.74 | 2.00 | 0.80 | 31.372 | 0.057 | 1.35 | 2.40 | NA | NA | NA | [17] |
| Pt | -0.01 | 0.75 | 2.99 | 1.08 | 14.756 | 0.134 | 1.75 | 2.69 | NA | NA | NA | [17] |
| Ti | -0.04 | 0.58 | 2.04 | 1.03 | 22.149 | 0.103 | 0.09 | 2.93 | NA | NA | NA | [17] |
| **$CrC_2N_4$** | | | | | | | | | | | | |
| G | 0.05 | 1.85 | 4.08 | 1.69 | 3.027 | 2.044 | 1.65 | 3.22 | NA | NA | NA | [17] |
| Ag | 0.07 | 1.84 | 3.52 | 1.51 | 5.487 | 1.493 | 1.10 | 3.15 | NA | NA | NA | [17] |
| Au | 0.32 | 1.54 | 4.10 | 1.55 | 4.012 | 1.598 | 1.36 | 3.18 | NA | NA | NA | [17] |
| Cu | -0.06 | 1.79 | 1.24 | 0.60 | 50.430 | 0.034 | 2.23 | 2.32 | NA | NA | NA | [17] |
| Ni | 0.15 | 1.81 | 0.32 | 0.13 | 92.742 | 0.000 | 1.17 | 2.05 | NA | NA | NA | [17] |
| Pd | 0.14 | 1.81 | 3.64 | 1.30 | 7.876 | 1.286 | 3.92 | 2.80 | NA | NA | NA | [17] |
| Pt | 0.19 | 1.73 | 3.69 | 1.22 | 9.060 | 1.328 | 3.29 | 2.77 | NA | NA | NA | [17] |
| Ti | -0.13 | 1.75 | 0.00 | 0.00 | 100 | 0.000 | 1.14 | 2.31 | NA | NA | NA | [17] |

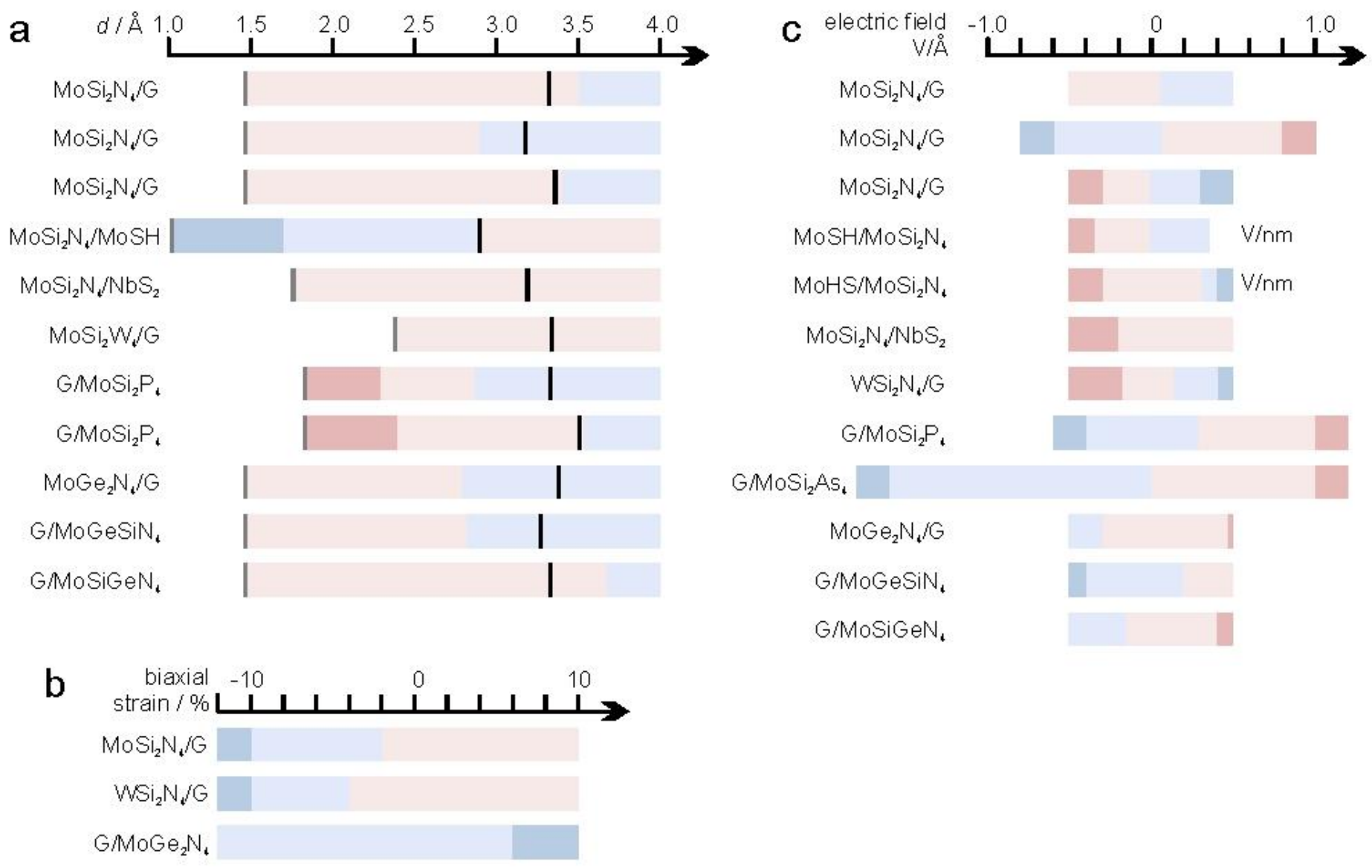

**Fig. S2**. Type of $MA_2Z_4$-metal structure as a function of **a**, vertical stress (interlayer distance), **b**, biaxial strain and **c**, electric field according to the data in **Table S1**. n-type and p-type are highlighted in blue and pink, Schottky and Ohmic - in weak and strong color, respectively. In **a**, the black vertical lines indicate the positions of the interlayer distance without vertical stress. The gray vertical lines indicate the positions of the minimal interlayer distance given by the sum of the covalent radii.

## 2. $MA_2Z_4$-semiconductor hetrostructures

**Table S2**. $MA_2Z_4$ - semiconductor (SC) heterostructures. BT - bandgap type (d - direct, i - indirect), BAL - band alignment type (I, II, III or h - hybrid). $E_g$ is the energy gap, ML - monolayer, H - heterostructure, EF - electric field, LM - lattice mismatch, *d* - interlayer distance.

| composition | BT | BAL | $E_g$ / eV PBE | $E_g$ / eV HSE | lattice constant ML / Å | lattice constant H / Å | LM % | *d* / Å | strain vertical | strain biaxial | EF | reference |
|---|---|---|---|---|---|---|---|---|---|---|---|---|
| **$MoSi_2N_4$** | | | | | | | | | | | | |
| BlueP/$MoSi_2N_4$ | i | II | 1.13 | 2.02 | 3.25/2.90 | 5.77 | 3 | 3.49 | | II to I at ε<-6%<br>II to I at ε>8% | to metal at EF<-0.5V/Å and EF>0.5V/Å<br>II to I at EF>0.3V/Å | [18] |
| $MoSi_2N_4$/BlueP | d | II | 1.175 | 2 | 2.90/3.27 | 5.8 | 2.4 | 3.49 | NA | ε=0-8%, shifted absorption spectra | NA | [19] |
| $MoSi_2N_4$/BP | d | I | 0.95 | NA | NA | 5.7 | 2.25 | 3.5 | NA | NA | NA | [11] |
| $C_2N$/$MoSi_2N_4$ | d | II | NA | 1.74 | 8.25/2.90 | 8.47 | <3 | NA | NA | NA | | [20] |
| $C_3N_4$/$MoSi_2N_4$ | d | II | 1.86 | 2.66 | 4.78/2.90 | 4.94 | 2.8 | 3.65 | II to I at *d*=2.65Å | II to I at ε<-2%<br>I to II at ε<-5%<br>II to I at ε>4%<br>d to i at ε>5% | II to I at EF<-0.4 V/Å | [21] |
| $MoSi_2N_4$/$CrCl_3$ (A stacking) | i | II | NA | NA | 2.91/5.932 | 5.8 | 0.29 | 3.198 | | NA | NA | [22] |
| $MoSi_2N_4$/$CrCl_3$ (B stacking) | i | II | NA | NA | 2.91/5.932 | 5.8 | 0.29 | 3.212 | NA | NA | NA | [22] |
| $MoSi_2N_4$/$CrS_2$ | i | II | 0.99 | 1.52 | 2.909/3.039 | 2.924 | 4.37 | 3.151 | remains type II | II to I at ε<-1.4% | NA | [23] |
| $MoSi_2N_4$/$Cs_3Bi_2I_9$ | d | II | 0.91 | NA | 2.90/8.65 | 8.66 | 0.227 | 3.55 | d to i at *d*<2.55Å | d to i at ε<-8% and ε>10%,<br>SC to metal at ε<-12% | NA | [24] |
| $MoSi_2N_4$/GaN | d | II | 1.7 | NA | NA | 5.7 | 2.28 | 3.3 | NA | NA | NA | [11] |
| $MoSi_2N_4$/GaN | d | I | 1.56 | NA | NA | NA | 2.68 | 3.44 | d,I to i,II at *d*<3.25Å | NA | d to i at EF<0<br>i to d at EF<-0.6V/Å | [25] |
| $MoSi_2N_4$/hBN | d | I | 1.66 | NA | NA | 5 | 0.14 | 3.3 | NA | NA | NA | [11] |
| $MoSi_2N_4$/InSe | d | II | 1.62 | NA | NA | 7.7 | 2.96 | 3.3 | NA | NA | NA | [11] |
| InSe/$MoSi_2N_4$ | d | II | 1.35 | 1.61(GGA-1/2) | 4.08/2.91 | 10.48 | 2.84 | 3.326 | NA | NA | NA | [26] |
| $MoSi_2N_4$/$MoGe_2N_4$ | i | II | 0.37 | NA | NA | 1.5 | 1.86 | 2.7 | NA | NA | NA | [11] |
| $MoGe_2N_4$/$MoSi_2N_4$ | i | II | 0.88-0.90 | 1.29-1.33 | NA/NA | 2.96 | NA | 3.33-3.35 | NA | NA | NA | [6] |
| $MoSi_2N_4$/$\alpha_2$-$MoGe_2P_4$ | d | II | 0.66 | NA | NA | 6 | 2.62 | 3.2 | NA | NA | NA | [11] |
| $MoSi_2N_4$/$MoS_2$ | i | II | 1.12 | NA | 2.91/3.18 | 2.96 | 4.43 | 3.11 | NA | II to I at ε<-3% and ε>1% | II to I at EF<-0.2V/Å<br>I to II at EF<-0.3V/Å<br>II to III at EF<-0.5V/Å and EF>0.4V/Å | [27] |
| $MoS_2$/$MoSi_2N_4$ | i | II | 1.26 | 1.84 | 3.21/2.91 | 2.96 | NA | 3.21 | NA | NA | NA | [28] |
| 2H-$MoSi_2N_4$/$MoS_2$ | i | II | NA | 2.08 | 2.902/3.18 | 2.96 | NA | NA | NA | NA | NA | [29] |
| 2H'-$MoSi_2N_4$/$MoS_2$ | i | II | NA | 1.93 | 2.91/3.18 | 2.96 | NA | NA | NA | NA | NA | [29] |
| $MoSi_2N_4$/2H-$MoS_2$ | d | h | 0.6 | NA | NA | 5.7 | 2.69 | 3.2 | NA | NA | NA | [11] |
| $MoSi_2N_4$/$MoSe_2$ | d | I | 1.37 | NA | NA | 5.8 | 0.58 | 3.3 | NA | NA | NA | [11] |
| $MoSe_2$/$MoSi_2N_4$ | d | I | 1.39 | 1.82 | 3.3/2.9 | 5.78 | 1.4 | 3.29 | NA | d to i at ε<-2%<br>I to II at ε<-7% and ε>1% | d,I to i,II at EF<-0.2 V/Å to d,II at EF>0.04 V/Å | [30] |
| 2H-$MoSi_2N_4$/MoSSe | i | II | NA | 1.26 | 2.902/3.25 | 2.97 | NA | NA | NA | NA | NA | [29] |
| 2H'-$MoSi_2N_4$/ MoSSe | i | I | NA | 1.55 | 2.91/3.25 | 2.98 | NA | NA | NA | NA | NA | [29] |
| $MoSi_2N_4$/$MoSi_2N_4$ | i | NA | 1.64 | 2.22 | 2.9/2.9 | 2.9 | 0 | NA | NA | i to d at ε<-4% | to metal at EF>6V/nm | [31] |
| $MoSi_2N_4$/$MoTe_2$ | i | I | 0.87 | NA | NA | 5.8 | 2.09 | 3.4 | NA | NA | NA | [11] |
| $MoSi_2N_4$/$Na_3Bi$ | NA | III | NA | NA | NA | 5.2 | 2.04 | 1.4 | NA | NA | NA | [11] |
| $MoSi_2N_4$/SMoSe | d | h | 0.88 | NA | NA | 5.7 | 1.64 | 3.2 | NA | NA | NA | [11] |
| $MoSi_2N_4$/SWSe | d | I | 1.13 | NA | NA | 5.7 | 1.61 | 3.2 | NA | NA | NA | [11] |
| $MoSi_2N_4$/SeMoS | d | I | 1.13 | NA | NA | 5.7 | 1.64 | 3.3 | NA | NA | NA | [11] |
| $MoSi_2N_4$/SeWS | d | I | 1.17 | NA | NA | 5.7 | 1.61 | 3.3 | NA | NA | NA | [11] |
| $MoSi_2N_4$/$TiSi_2N_4$ | i | NA | 0 | 0.343 | NA/NA | 5.841 | NA | NA | NA | to metal at ε<-4% | | [32] |
| $MoSi_2N_4$/$WGe_2N_4$ | i | II | 0.99 | NA | NA | 1.5 | 1.89 | 2.8 | NA | NA | NA | [11] |
| $MoSi_2N_4$/$\alpha_2$-$WGe_2P_4$ | d | h | 0.78 | NA | NA | 6 | 2.1 | 3.2 | NA | NA | NA | [11] |
| $MoSi_2N_4$/2H-$WS_2$ | d | I | 1.05 | NA | NA | 5.7 | 2.64 | 3.2 | NA | NA | NA | [11] |
| $MoSi_2N_4$/$WSe_2$ | i | II | 1.4 | NA | NA | 5.8 | 1.09 | 3.3 | NA | NA | NA | [11] |
| $WSe_2$/$MoSi_2N_4$ | i | II | 1.38 | 1.81 | 3.3/2.90 | 5.8 | NA | 3.25 | NA | NA | NA | [33] |
| $WSe_2$/$MoSi_2N_4$ | d | I | 1.17 | 1.57 | 3.29/2.9 | 5.78 | 1.4 | 3.29 | NA | NA | d,I to I,II at EF<-0.05V/Å; to d,II at EF>0.4 V/Å | [34] |
| $MoSi_2N_4$/$WSi_2N_4$ | d | II | 1.36 | NA | NA | 1.4 | 0.05 | 3 | NA | NA | NA | [11] |
| $MoSi_2N_4$/$WSi_2N_4$ | i | NA | 1.92 | 2.35 | NA/NA | 5.824 | NA | NA | NA | i to d at ε<-3% | | [32] |
| $WSi_2N_4$/$MoSi_2N_4$ | i | II | NA | 1.9 | 2.915/2.909 | 2.902 | ~0 | 3.01 | NA | NA | NA | [35] |
| $MoSi_2N_4$/$WTe_2$ | i | I | 1.09 | NA | NA | 5.8 | 2.18 | 3.4 | NA | NA | NA | [11] |
| $MoSi_2N_4$/ZnO | i | II | 1.61 | NA | NA | 5.8 | 1.04 | 3.6 | NA | NA | NA | [11] |
| $MoSi_2N_4$/ZnO | i | II | 1.6 | NA | NA | NA | 0.92 | 3.15 | i,II to d,I at *d*>3.7Å | NA | i to d at EF>0.2V/Å<br>i to d at EF<-0.6V/Å | [25] |
| **$WSi_2N_4$** | | | | | | | | | | | | |
| $WSi_2N_4$/BP | d | I | 0.96 | NA | NA | 5.7 | 2.3 | 3.5 | NA | NA | NA | [11] |
| $WSi_2N_4$/GaN | d | I | 1.84 | NA | NA | 5.7 | 2.34 | 3.3 | NA | NA | NA | [11] |
| $WSi_2N_4$/hBN | d | I | 2.04 | NA | NA | 5 | 0.2 | 3.3 | NA | NA | NA | [11] |
| $WSi_2N_4$/InSe | d | II | 1.43 | NA | NA | 7.7 | 2.91 | 3.3 | NA | NA | NA | [11] |
| $WSi_2N_4$/$MoGe_2N_4$ | i | II | 0.25 | NA | NA | 1.5 | 1.8 | 2.8 | NA | NA | NA | [11] |
| $WSi_2N_4$/$\alpha_2$-$MoGe_2P_4$ | d | II | 0.44 | NA | NA | 6 | 2.57 | 3.2 | NA | NA | NA | [11] |
| $WSi_2N_4$/2H-$MoS_2$ | d | h | 0.81 | NA | NA | 5.7 | 2.74 | 3.2 | NA | NA | NA | [11] |
| $MoSe_2$/$WSi_2N_4$ | d | I | 1.26 | NA | NA | NA | 0.82 | 3.25 | I to II at *d*>4Å | I to II at $\varepsilon$>1%<br>d to i at $\varepsilon\leq$-2% | I to II at EF<-0.025V/Å and EF>0.075V/Å<br>II to III at EF<-0.125V/Å and EF>0.2V/Å | [36] |
| $WSi_2N_4$/$MoSe_2$ | d | II | 1.11 | NA | NA | 5.8 | 0.63 | 3.3 | NA | NA | NA | [11] |
| $WSi_2N_4$/$MoTe_2$ | i | I | 0.89 | NA | NA | 5.8 | 2.04 | 3.5 | NA | NA | NA | [11] |
| $WSi_2N_4$/$Na_3Bi$ | NA | III | NA | NA | NA | 5.2 | 1.98 | 1.4 | NA | NA | NA | [11] |
| $WSi_2N_4$/SMoSe | d | II | 0.73 | NA | NA | 5.7 | 1.69 | 3.2 | NA | NA | NA | [11] |
| $WSi_2N_4$/SeMoS | d | I | 1.11 | NA | NA | 5.7 | 1.69 | 3.3 | NA | NA | NA | [11] |
| $WSi_2N_4$/SeWS | d | I | 1.28 | NA | NA | 5.7 | 1.67 | 3.3 | NA | NA | NA | [11] |
| $WSi_2N_4$/SWSe | d | h | 1.02 | NA | NA | 5.7 | 1.67 | 3.2 | NA | NA | NA | [11] |
| $WSi_2N_4$/$WGe_2N_4$ | i | II | 0.78 | NA | NA | 1.5 | 1.84 | 2.8 | NA | NA | NA | [11] |
| $WSi_2N_4$/$\alpha_2$-$WGe_2P_4$ | d | II | 0.59 | NA | NA | 6 | 2.05 | 3.2 | NA | NA | NA | [11] |
| $WSi_2N_4$/2H-$WS_2$ | d | I | 1.04 | NA | NA | 5.7 | 2.69 | 3.2 | NA | NA | NA | [11] |
| $WSi_2N_4$/$WSe_2$ | d | I | 1.45 | NA | NA | 5.8 | 1.67 | 3.3 | NA | NA | NA | [11] |
| $WSe_2$/$WSi_2N_4$ | d | I | 1.16 | NA | NA | NA | 0.64 | 3.27 | remains type I | I to II at $\varepsilon$>3%<br>d to i at $\varepsilon\leq$-2% | I to II at EF<-0.05V/Å and EF>0.05V/Å<br>II to III at EF<-0.2V/Å and EF>0.175V/Å | [36] |
| $WSi_2N_4$/$WSi_2N_4$ | i | NA | 1.94 | 2.56 | 2.9/2.9 | 2.9 | 0 | NA | NA | NA | to metal at EF>6V/nm | [31] |
| $WSi_2N_4$/$WTe_2$ | i | I | 1.19 | NA | NA | 6 | 2.13 | 3.5 | NA | NA | NA | [11] |
| $WSi_2N_4$/ZnO | d | I | 1.83 | NA | NA | 5.8 | 1.1 | 3.1 | NA | NA | NA | [11] |
| | | | | | | | | | | | | |
| BP/$MoSi_2P_4$ | d | II | NA | 1.02 | 3.21/3.47 | NA | 3.9 | 3.39 | NA | NA | NA | [37] |
| BP/$MoGe_2P_4$ | d | II | 0.34-0.47 | 0.62-0.77 | 3.2/3.04 | 3.12 | 2.56 | 3.08-3.38 | II to I at *d*<2.7Å | NA | II to I at EF>0.15 V/Å | [38] |
| $(MoSi_2N_4)_5$/$(MoSiGeN_4)_0$ | i | NA | 1.925 | 2.385 | NA/2.96 | 2.91 | NA | NA | NA | NA | NA | [39] |
| $(MoSi_2N_4)_4$/$(MoSiGeN_4)_1$ | i | NA | 1.823 | 2.71 | NA/2.96 | NA | NA | NA | NA | NA | NA | [39] |
| $(MoSi_2N_4)_3$/$(MoSiGeN)_2$ | i | NA | 1.716 | 2.137 | NA/2.96 | NA | NA | NA | NA | NA | NA | [39] |
| $(MoSi_2N_4)_2$/$(MoSiGeN_4)_3$ | i | NA | 1.598 | 2.021 | NA/2.96 | 2.93 | NA | NA | NA | NA | NA | [39] |
| $(MoSi_2N_4)_1$/$(MoSiGeN_4)_4$ | i | NA | 1.494 | 1.979 | NA/2.96 | NA | NA | NA | NA | NA | NA | [39] |
| $(MoSi_2N_4)_0$/$(MoSiGeN_4)_5$ | i | NA | 1.385 | 1.86 | NA/2.96 | 2.9 | NA | NA | NA | NA | NA | [39] |
| $T_4$-MoSTe/$MoGe_2N_4$ | NA | NA | 0 | NA | 3.362/3.037 | NA | 1.48 | 3.4 | to i gap SC at *d*>3.63Å | semi-metal so SC at $\varepsilon$<–2% | NA | [40] |
| $S_4$-MoSTe/$MoGe_2N_4$ | d | II | 0.27 | NA | 3.362/3.037 | NA | 1.48 | 3.1 | no gap at *d*<2.3Å | SC to semi-metal at $\varepsilon$>2% | NA | [40] |
| MoSSe/$MoGeSiN_4$ | d | II | 0.87 | 1.28 | 3.25/2.96 | 8.98 | 3.3 | 3.29 | | d to i at $\varepsilon$<-2%<br>no gap at $\varepsilon$>8% | NA | [41] |

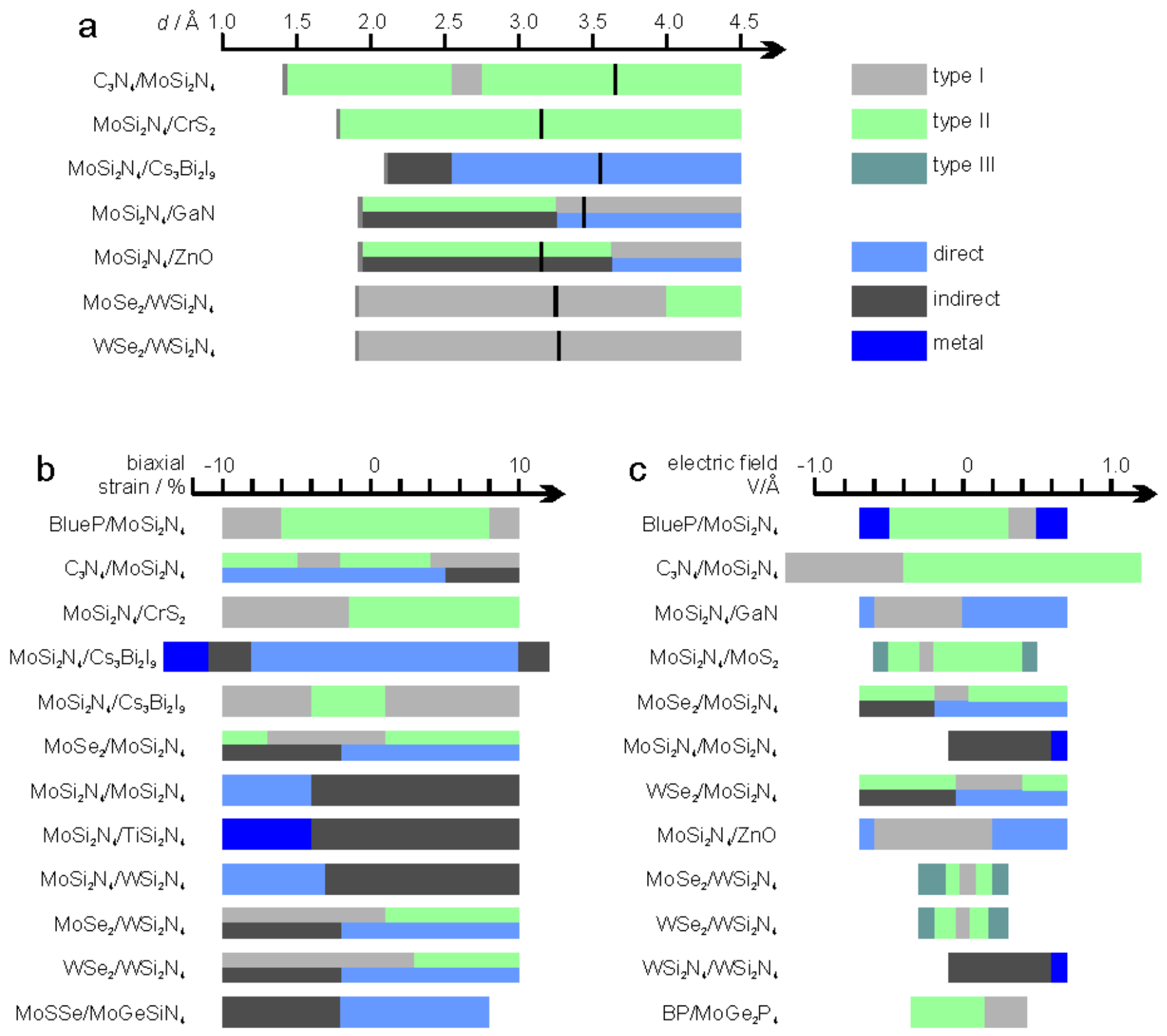

**Fig. S3**. Type of $MA_2Z_4$-heterostructure as a function of **a**, vertical stress (interlayer distance), **b**, biaxial strain and **c**, electric field according to the data shown in **Table S2**. In **a**, the black vertical lines indicate the positions of the interlayer distance without vertical stress. The gray vertical lines indicate the positions of the minimal interlayer distance given by the sum of the covalent radii.

## 3. Catalytic properties of $MA_2Z_4$ materials

Reaction potential depends on pH as defined by the Nernst equation: $E^{o}$(pH) = $E^{o}$(pH=0) + pH·0.059 eV, (at 25° C). The band edges of a semiconductor can be evaluated either directly from its energy bands distribution, or from its bandgap $E_g$ alone by using the following equations [42] (Liu2011APL): $E_{\mathrm{VBE}} = x - 4.5\mathrm{eV} + 0.5E_{\mathrm{g}}$, and $E_{\mathrm{CBE}} = E_{\mathrm{VBE}} - E_{\mathrm{g}}$, where $x$ is the geometric average value of the electronegativity of the constituent atoms ($x$ = 5.195 for $MoSi_2N_4$ [43] (Bartolotti1980JACS)), and 4.5 eV is the energy of free electrons on the hydrogen scale.

The water splitting reaction $\mathrm{H_2O} \rightarrow \mathrm{H_2} + 1/2\mathrm{O_2}$ consists of two half-reactions [44] (Nishioka2023NatRevMethods): hydrogen evolution reaction (HER) $2\mathrm{H}^+ + 2\mathrm{e}^- \rightarrow \mathrm{H_2}$ and oxygen evolution reaction (OER) $\mathrm{H_2O} + 2\mathrm{h}^+ \rightarrow 1/2\mathrm{O_2} + 2\mathrm{H}^+$. Thus, the catalyst's VBE should be lower than the potential of the OER ($E(H_2O/O_2)$ =-5.67 eV + pH·0.059 eV), its CBE should be higher than the potential of HER ($E(H^+/H_2)$ =-4.44 eV + pH·0.059 eV), with the minimal energy gap of 1.23 eV. Photocatalytic $CO_2$ reduction reaction can be realized by one of the following reactions:

$CO_2 + 2H^+ + 2e \rightarrow HCOOH$ ($E°$=-0.194 V) (1)

$CO_2 + 2H^+ + 2e \rightarrow CO + H_2O$ ($E°$=-0.099 V) (2)

$CO_2 + 6H^+ + 6e \rightarrow CH_3OH + 2H_2O$ ($E°$=0.031 V) (3)

$CO_2 + 8H^+ + 8e \rightarrow CH_4 + 2H_2O$ ($E°$=0.171 V) (4)

The potentials $E°$ shown here (vs standard hydrogen electrode (SHE) at pH=0) were calculated using the free energy [45] (Latimer_book). The positions of the potentials for water splitting and $CO_2$ reduction reactions, together with the band edges of a hypothetically suitable catalyst are shown in Fig. 7a.

Table S3. Catalytic properties of $MA_2Z_4$ materials. BT - bandgap type (d - direct, i - indirect), $x$ - the geometric average of the electronegativity of the constituent atoms. WS - photo-catalytic water splitting, $CO_2$ - $CO_2$ reduction reactions, where "1,2,3,4" is the reaction number as defined above, $\alpha$ - absorption coefficient, given for visible light range: 1.59 - 3.26 eV (780 - 380nm). ML - monolayer, SAC - single atom catalyst, and H - hetero-structures.

| composition | BT | $E_g$ / eV PBE | $E_g$ / eV HSE | $x$ | $E_{CBM}$ /eV NHE | $E_{VBM}$ / eV NHE | WS | $CO_2$ | $\alpha$ $10^5$ $cm^{-1}$ | strain biaxial | reference |
|---|---|---|---|---|---|---|---|---|---|---|---|
| **ML** | | | | | | | | | | | |
| $CrSi_2N_4$ | i | 0.49 | 0.99 | NA | 0.86 | 1.85 | no | no | NA | NA | [46] |
| $HfSi_2N_4$ | i,d | 1.77(i) | 2.84(d) | NA | -0.14 | 2.7 | yes | 2,3,4 | 0.3-1.8 | NA | [46] |
| $MoGe_2N_4$ | I | 0.85 | 1.42 | NA | 0.61 | 2.03 | no | no | 0.6-4.0 | NA | [46] |
| 2H-$MoGe_2N_4$ | i | 0.91 | 1.27 | NA | 0.46 | 1.86 | no | no | 0.3-2.0 | | [47] |
| $MoSi_2N_4$ | i | 1.72 | 2.21 | NA | 0.28 | 2.49 | no | no | NA | NA | [46] |
| $MoSi_2N_4$ | i | 1.38 | 2.13 | NA | -1.06 | 1.11 | no | no | NA | $\varepsilon$=2% shifts band edges to CBM=-4.12(-0.32)eV and VBM=-5.85(1.41)eV so that WS and $CO_2$ reduction (1,2,3,4) are possible | [48] |
| 2H-$MoSi_2N_4$ | i | 1.79 | 2.23 | NA | -0.64 | 1.66 | yes | 1,2,3,4 | 0.2-2.9 | | [47] |
| $TiSi_2N_4$ | i | 1.61 | 2.74 | NA | -0.12 | 2.62 | yes | 2,3,4 | NA | NA | [46] |
| $WGe_2N_4$ | i | 1.18 | 1.6 | NA | 0.61 | 2.21 | no | no | NA | NA | [46] |
| 2H-$WGe_2N_4$ | i,d | 1.15(i) | 1.51(d) | NA | -0.14 | 1.41 | yes | 2,3,4 | 0.2-1.7 | | [47] |
| $WSi_2N_4$ | i | 2.19 | 2.51 | NA | 0.19 | 2.7 | no | no | NA | NA | [46] |
| $WSi_2N_4$ | NA | 1.65 | 2.38 | NA | -1.47 | 0.92 | no | no | NA | $\varepsilon$=3% shifts band edges to: CBM=-4.09(-0.35)eV and VBM=-5.76(1.32)eV<br>$\varepsilon$=4% shifts band edges to: CBM=-4.439(-0.01)eV and VBM=-5.91(1.47)eV<br>so that WS and $CO_2$ reduction (1,2,3,4) are possible | [48] |
| 2H-$WSi_2N_4$ | i | 2.07 | 2.57 | NA | -1.04 | 1.26 | yes | 1,2,3,4 | 0.2-0.6 | | [47] |
| $ZrSi_2N_4$ | i,d | 1.56(i) | 2.78(d) | NA | -0.08 | 2.7 | yes | 3,4 | NA | NA | [46] |
| **ML, Janus** | | | | | | | | | | | |
| $MoSi_2N_4$ | i | NA | 2.32 | NA | -0.81 | 1.51 | yes | 1,2,3,4 | 0.1-0.6 | uniaxial: both band edges position increase as a function of $\varepsilon$ ($\varepsilon$ increasing, band edges decreasing) | [49] |
| $MoSiGeN_4$ | i | NA | 1.81 | NA | -0.55 | 1.86 | yes | 1,2,3,4 | 0.1-3.6 | uniaxial: both band edges position increase as a function of $\varepsilon$ ($\varepsilon$ increasing, band edges decreasing) | [49] |
| $WSiGeN_4$ | i | NA | 2.25 | NA | -1.28 | 1.56 | yes | 1,2,3,4 | 0.1-0.6 | uniaxial: both band edges position increase as a function of $\varepsilon$ ($\varepsilon$ increaseing, band edges decreasing) | [49] |
| **ML, strain** | | | | | | | | | | | |
| $MoSi_2N_4$ | i | 1.96 | NA | NA | -0.71 | 1.42 | yes | 1,2,3,4 | 0-0.2 | | [50] |
| $MoSi_2N_4$ $\varepsilon$=6% | i | 0.91 | NA | NA | NA | NA | no | no | 0.5-5.0 | $\varepsilon$=6% shifts abs spectrum by 100 nm to longer wavelength | [50] |
| $MoSi_2N_4$ $\varepsilon$=-6% | i | 2.76 | NA | NA | -1.32 | 1.88 | yes | 1,2,3,4 | 0-3.0 | $\varepsilon$=-6% shifts abs spectrum by 100 nm to shorter wavelength | [50] |
| **SAC** | | | | | | | | | | | |
| $MoSi_2N_4$ | i | 1.89 | NA | 5.195 | -0.250 | 1.640 | yes | 1,2,3,4 | 0.2-3.0 | NA | [51] |
| 1Au | i | 1.81 | NA | NA | -0.210 | 1.600 | yes | 1,2,3,4 | 0.4-3.1 | NA | [51] |
| 2Au | i | 1.78 | NA | NA | -0.195 | 1.585 | yes | 1,2,3,4 | NA | NA | [51] |
| 3Au | i | 1.42 | NA | NA | -0.015 | 1.405 | yes | 3,4 | 0.4-3.4 | NA | [51] |
| 5Au | i | 0.2 | NA | NA | 0.595 | 0.795 | no | no | NA | NA | [51] |
| 7Au | i | 0.07 | NA | NA | 0.660 | 0.730 | no | no | 0.06-0.51 | NA | [51] |
| 9Au | i | 0 | NA | NA | 0.695 | 0.695 | no | no | 0.18-0.57 | NA | [51] |
| **SAC** | | | | | | | | | | | |
| $MoSi_2N_4$ | i | 1.89 | NA | 5.195 | -0.25 | 1.64 | yes | 1,2,3,4 | 0.02-0.34 | NA | [52] |
| 1Li | i | 1.73 | NA | 5.139 | -0.226 | 1.504 | yes | 1,2,3,4 | 0.57-0.92 | NA | [52] |
| 2Li | i | 1.58 | NA | 5.085 | -0.205 | 1.375 | yes | 1,2,3,4 | NA | NA | [52] |
| 3Li | i | 1.42 | NA | 5.033 | -0.177 | 1.243 | yes | 2,3,4 | NA | NA | [52] |
| 1Na | i | 1.61 | NA | 5.13 | -0.175 | 1.435 | yes | 2,3,4 | 1.09-1.34 | NA | [52] |
| 2Na | i | 1.53 | NA | 5.068 | -0.197 | 1.333 | yes | 1,2,3,4 | NA | NA | [52] |
| 3Na | i | 1.46 | NA | 5.008 | -0.222 | 1.238 | yes | 1,2,3,4 | NA | NA | [52] |
| 1K | i | 1.75 | NA | 5.115 | -0.26 | 1.49 | yes | 1,2,3,4 | 0.89-1.14 | NA | [52] |
| 2K | i | 1.69 | NA | 5.039 | -0.306 | 1.384 | yes | 1,2,3,4 | NA | NA | [52] |
| 3K | i | 1.47 | NA | 4.966 | -0.269 | 1.201 | yes | 1,2,3,4 | NA | NA | [52] |
| **H** | | | | | | | | | | | |
| $MoSi_2N_4$ | i | 1.78 | 2.36 | NA | -0.82 | 1.54 | yes | 1,2,3,4 | 0.16-0.8 | NA | [19] |
| BlueP | i | 1.94 | 2.76 | NA | -0.46 | 2.3 | yes | 1,2,3,4 | 0.14-0.5 | NA | [19] |
| $MoSi_2N_4$/ BlueP | i | 1.175 | 2 | NA | -0.71 | 1.63 | yes | 1,2,3,4 | 0.16-2.0 | $\varepsilon$=0 to (+6%) shifts absorption spectrum by 0 to (-1)eV; $\varepsilon$=-0-(-8%) by 0 to (+1eV) | [19] |
| | | | | | | | | | | | |
| $MoSi_2N_4$ | i | 1.78 | 2.31 | NA | -0.77 | 1.58 | yes | 1,2,3,4 | 0-1.0 | NA | [53] |
| AB BL $MoSi_2N_4$ | i | 1.7 | 2.23 | NA | -0.7 | 1.23 | yes | 1,2,3,4 | 0-3.5 | | [53] |
| | | | | | | | | | | | |
| 2H-$MoSi_2N_4$/ $MoS_2$ | i | NA | 2.08 | NA | NA | NA | yes | 1,2,3,4 | NA | NA | [29] |
| 2H'-$MoSi_2N_4$/ $MoS_2$ | i | NA | 1.93 | NA | NA | NA | yes | 1,2,3,4 | NA | NA | [29] |
| 2H-$MoSi_2N_4$/ MoSSe | i | NA | 1.26 | NA | NA | NA | no | no | NA | NA | [29] |
| 2H'-$MoSi_2N_4$/ MoSSe | i | NA | 1.55 | NA | NA | NA | no | no | NA | NA | [29] |
| | | | | | | | | | | | |
| $WSe_2$/ $MoSi_2N_4$ | i, II | 1.38 | 1.81 | NA | -0.48 | 1.33 | yes | 1,2,3,4 | 0.3-2.2 | NA | [33] |
| InSe/ $MoSi_2N_4$ | d, II | 1.35 | 1.61(GGA-1/2) | NA | -0.18 | 1.43 | yes | 2,3,4 | 0-1.5 | NA | [26] |
| $C_2N/MoSi_2N_4$ | d, II | NA | 1.74 | NA | -0.2 | 1.54 | yes | 1,2,3,4 | 0-4 | NA | [20] |
| $MoSi_2N_4/CrS_2$ | i, II | 0.99 | 1.52 | NA | NA | NA | yes | NA | NA | | [23] |
| $WSi_2N_4/$ $MoSi_2N_4$ | i, II | NA | 1.9 | NA | -0.69 | 1.3 | yes | NA | 0.2-3.2 | NA | [35] |
| | | | | | | | | | | | |
| $(MoSi_2N_4)_5/$ $(MoSiGeN_4)_0$ | | 1.925 | 2.385 | NA | -5.463 | 1.660 | yes | 1,2,3,4 | 0-0.35 | NA | [39] |
| $(MoSi_2N_4)_4/$ $(MoSiGeN_4)_1$ | | 1.823 | 2.71 | NA | NA | NA | yes | 1,2,3,4 | NA | NA | [39] |
| $(MoSi_2N_4)_3/$ $(MoSiGeN)_2$ | | 1.716 | 2.137 | NA | NA | NA | yes | 1,2,3,4 | NA | NA | [39] |
| $(MoSi_2N_4)_2/$ $(MoSiGeN_4)_3$ | | 1.598 | 2.021 | NA | NA | NA | yes | 2,3,4 | 0-0.35 | NA | [39] |
| $(MoSi_2N_4)_1/$ $(MoSiGeN_4)_4$ | | 1.494 | 1.979 | NA | NA | NA | yes | 3,4 | NA | NA | [39] |